\documentclass[aoas]{imsart}

\RequirePackage{amsthm,amsmath,amsfonts,amssymb}
\RequirePackage[authoryear]{natbib}
\RequirePackage[colorlinks,citecolor=blue,urlcolor=blue]{hyperref}
\RequirePackage{graphicx}

\usepackage{enumerate}
\usepackage{url} 
\usepackage{soul}
\usepackage{xcolor}
\usepackage{amsfonts}
\usepackage{algorithm}
\usepackage{algorithmic}
\usepackage{multirow}
\usepackage{bm}
\usepackage{lscape}
\usepackage{ulem}
\usepackage{hyperref}
\usepackage{rotating}
\usepackage{adjustbox}
\usepackage{longtable}
\usepackage{makecell}

\newcommand{\E}{\operatorname{E}}

\newcommand{\I}{\operatorname{I}}
\newcommand{\N}{\operatorname{N}}

\newcommand{\U}{\operatorname{U}}

\usepackage{amsmath}

\DeclareMathOperator*{\argmin}{argmin}

\startlocaldefs
\theoremstyle{plain}

\theoremstyle{remark}

\endlocaldefs

\begin{document}

\begin{frontmatter}
\title{RZiMM-scRNA: A regularized zero-inflated mixture model framework for single-cell RNA-seq data}
\runtitle{RZiMM-scRNA}

\begin{aug}
\author[A]{\fnms{Xinlei} \snm{Mi}\ead[label=e1]{xinlei.mi@northwestern.edu}},
\author[B]{\fnms{William} \snm{Bekerman}\ead[label=e2,mark]{wjb239@cornell.edu}}
\author[E]{\fnms{Peter A.} \snm{Sims}\ead[label=e5,mark]{pas2182@cumc.columbia.edu}}
\author[C]{\fnms{Peter D.} \snm{Canoll}\ead[label=e3,mark]{pc561@cumc.columbia.edu}}
\and
\author[D]{\fnms{Jianhua} \snm{Hu}\ead[label=e4,mark]{jh3992@cumc.columbia.edu}}
\address[A]{Department of Preventive Medicine, Northwestern University Feinberg School of Medicine, 
\printead{e1}}

\address[B]{Department of Statistics and Data Science, Cornell University, 
\printead{e2}}

\address[E]{Department of Systems Biology, Columbia University, 
\printead{e5}}

\address[C]{Department of Pathology and Cell Biology, Columbia University, 
\printead{e3}}

\address[D]{Department of Biostatistics, Columbia University, 
\printead{e4}}
\end{aug}

\begin{abstract}
Applications of single-cell RNA sequencing in various biomedical research areas have been blooming. This new technology provides unprecedented opportunities to study disease heterogeneity at the cellular level. However, unique characteristics of scRNA-seq data, including large dimensionality, high dropout rates, and possibly batch effects, bring great difficulty into the analysis of such data. Not appropriately addressing these issues obstructs true scientific discovery. Herein, we propose a unified Regularized Zero-inflated Mixture Model framework designed for scRNA-seq data (RZiMM-scRNA) to simultaneously detect cell subgroups and identify gene differential expression based on a developed importance score, accounting for both dropouts and batch effects.  
We conduct extensive simulation studies in which we evaluate the performance of RZiMM-scRNA and compare it with several popular methods, including Seurat, SC3, K-Means, and Hierarchical Clustering. Simulation results show that RZiMM-scRNA demonstrates superior clustering performance and enhanced biomarker detection accuracy compared to alternative methods, especially when cell subgroups are less distinct, verifying the robustness of our method.

\textcolor{black}{Our empirical investigations focus on two brain tumor studies dealing with astrocytoma of various grades, including the most malignant of all brain tumors, glioblastoma multiforme (GBM). Our goal is to delineate cell heterogeneity and identify driving biomarkers associated with these tumors. Notably, RZiMM-scNRA successfully identifies a small group of oligodendrocyte cells which has drawn much attention in biomedical literature on brain cancers. In addition, our method discovers several new biomarkers which are not discussed in the original studies, including PLP1, BCAN, and PTPRZ1 --- all associated with the development and malignant growth of glioma --- as well as CAMK2B, which is downregulated in glioma and GBM, and implicated in neurodevelopment, brain function, learning and memory processes.
     }
    
\end{abstract}

\begin{keyword}
\kwd{Single cell}
\kwd{Mixture model}
\kwd{Clustering}
\kwd{Batch effect}
\kwd{Dropout}
\end{keyword}

\end{frontmatter}

\section{Introduction}

Single-cell RNA sequencing (scRNA-seq) techniques have recently emerged and enabled large volumes of transcriptome profiling of individual cells \citep{wang2010single, prakadan2017scaling}. Unlike conventional gene analyses of bulk samples where an average of gene expression levels in a bulk cell population is measured and individual cell characteristics are ignored, scRNA-seq data provide detailed information of gene expressions at the level of individual cells \citep{hedlund2018single}. Research using scRNA-seq has led to detection of new cell types, delineation of cell heterogeneity, identification of gene regulatory mechanisms, cell developmental dynamics, and other important discoveries \citep{xue2013genetic, treutlein2014reconstructing, grun2015single, petropoulos2016single, reinius2016analysis}. 

\textcolor{black}{Our empirical investigations aim to advance understanding of brain tumors, including glioblastoma multiforme (GBM), which is fast-growing and the most aggressive of all brain tumors. The first study we encounter focuses on various grades of astrocytoma, the most common type of primary brain tumors affecting adults and can occur in most parts of the brain and in the spinal cord. These tumors are quite resistant to therapies and often progress rapidly to high-grade lesions \citep{de2017expression}. The astrocytoma scRNA-seq data is generated from ten IDH-mutant human astrocytoma tumors of different grades by \citeauthor{venteicher2017decoupling}, and available in the Gene Expression Omnibus (GEO) database under accession number GSE89567 \citep{barrett2012ncbi}.  In the second study, high-grade astrocytoma scRNA-seq count data is generated by \citeauthor{yuan2018single} and available in the GEO database under accession number GSE103224. In this study, samples from eight patients were taken of the cellular milieu of high-grade gliomas, namely the most aggressive types --- GBM and anaplastic astrocytoma. We intend to identify hidden cell subgroup structures and discover important biomarkers which drive these structures.}

Despite its great promise in advancing our knowledge of diseases, several challenges are encountered in the analysis of such scRNA-seq data, prompting the development of novel analytical strategies  \citep{kiselev2019challenges, petegrosso2019machine}. 
Following quality control and normalization, an essential aspect of scRNA-seq data analysis is unsupervised clustering to identify hidden cell subgroup structures and the biomarkers which drive these patterns. However, many problems arise when clustering scRNA-seq data \citep{kiselev2019challenges}, such as high dimensionality, high rate of dropouts, and the presence of sample/batch effects. High dimensionality refers to the large numbers of cells and genes in a tissue sample. Dropout events are zero expressions of transcripts in single cells, due to low expressions/undetectable levels, false quantification, or failure in amplification \citep{stegle2015computational, kiselev2019challenges, petegrosso2019machine}, which can not be addressed via simple normalization \citep{vallejos2017normalizing}. Sample/batch effects may be induced by sample heterogeneity, or technical issues and experimental designs \citep{buttner2017assessment}. Failure to appropriately account for these effects can introduce non-trivial bias and significantly impact clustering algorithms \citep{tung2017batch}. 

On the other hand, many methods have been developed to particularly deal with dropouts based on imputation techniques and zero-inflated models. For example, CIDR \citep{lin2017cidr}, CMF-Impute \citep{xu2020cmf} and ZIFA \citep{pierson2015zifa} impute gene expression values in dropout events based on cell and gene similarities or parametric modeling with latent variables. However, imputation methods usually require various conditions and may lead to new bias and the elimination of biologically meaningful variation. SC3 \citep{kiselev2017sc3} is a popular pipeline for scRNA-seq data using combined clustering methods, while it does not perform any form of batch effect correction. Existing approaches to correct for batch effects include model-based adjustment, such as Combat \citep{johnson2007adjusting}; canonical correlation analysis (CCA) subspace alignment, such as Seurat \citep{butler2018integrating}; and mutual nearest neighbors (MNN), such as fastMNN \citep{haghverdi2018batch} and BEER \citep{zhang2019novel}. 
Unfortunately, these existing methods have various drawbacks. For example, Combat \citep{johnson2007adjusting} was initially designed for bulk cell data, and thus provides no special treatment for dropout events. Seurat \citep{butler2018integrating}, fastMNN \citep{haghverdi2018batch}, BEER \citep{zhang2019novel}, and ZINB-WaVE \citep{risso2018general} first perform dimensionality reduction and correct for batch effects, followed by cell clustering. However, these multi-staged approaches often lose information across the separate steps, as the analysis conducted at a given step is conditional on the summarized data obtained in the previous stage, and therefore cannot recover any useful information that might have been lost then. In summary, none of these discussed methods provide a unified strategy to systematically analyze scRNA-seq data.

We also remark that the most common practice in scRNA-seq data analysis is still to analyze a single biological sample at a time, followed by manually summarizing the common findings shared by different samples. Yet this strategy fails to leverage information shared across samples and potentially sacrifices the detection power. To address this drawback, our method will take into account all the samples jointly in the new modeling system. 

Specifically, we propose RZiMM-scRNA, a Regularized Zero-inflated Mixture Model framework to simultaneously cluster cells and identify differentially expressed genes between cell subgroups in scRNA-seq data. RZiMM-scRNA also accounts for dropout events and sample heterogeneity. In particular, estimation of the latent group allocation and effect is directly linked to cell clustering membership construction. An $l_1$-penalty on the pairwise differences of group effects within a gene is also incorporated to impose sparsity on the space of differentially expressed genes as a treatment for high dimensionality. Lastly, the parametric nature of RZiMM-scRNA naturally allows for the development of an importance score for biomarker identification, whose statistical significance can also be assessed. Unlike popular two-staged methods which are unable to fully utilize information in the data, RZiMM-scRNA is a unified model framework that addresses all of the discussed challenges. We demonstrate the promise of the proposed method via extensive simulation studies and applications to two brain cancer scRNA-seq datasets.

\section{Method}
\subsection{A New Model Framework}
We start with the notations.
Let $X_{ij}$ be the normalized expression of the $j^\text{th}$ gene from the $i^\text{th}$ cell, for $i=1,...,N$, $j=1,...,J$. Note that the cells from all the samples (patients) are collapsed together in our analysis. To account for sparse values in the single-cell gene expression data, we assume that $X_{ij}$ follows a zero-inflated mixture distribution,
\begin{eqnarray}
\operatorname{P}(X_{ij}=x|\mu_{ij},\sigma_j^2) = (1-\pi_{j})\I(x=0) + \pi_{j}\I(x\ne0)\cdot\mathcal{G}(x|
\mu_{ij}, \sigma_j^2), 
\label{assump0}
\end{eqnarray}
where $\pi_j$ is the probability that gene $j$ has a non-zero value and $\mathcal{G}(\cdot|
\mu_{ij}, \sigma_j^2)$ is a specified distribution for  non-zero values of gene $j$, with mean $\mu_{ij}$ in cell $i$ and variance $\sigma_j^2$. The proposed framework can accommodate a wide range of data types by specifying $\mathcal{G}(\cdot| \mu_{ij}, \sigma_j^2)$ with different forms of distributions, including Gaussian, Poisson, and negative binomial, which have been widely used in modeling scRNA-seq data. 

\begin{figure}
    \centering
    \includegraphics[width = 0.7\textwidth]{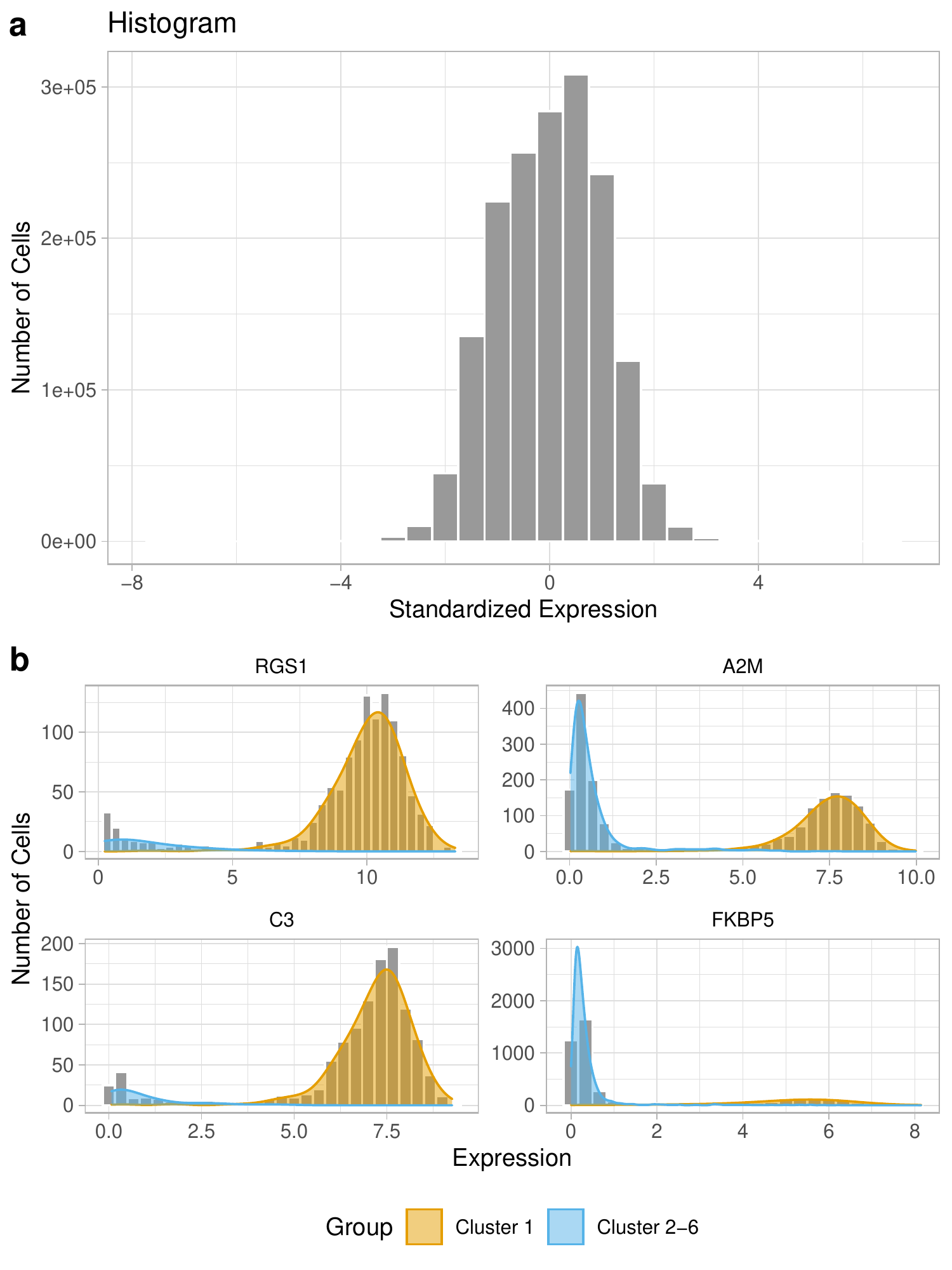}
    \caption{Distribution of Expression in Astrocytoma scRNA-seq Data. a. Histogram of standardized non-zero expression values from five hundred randomly selected genes. b. Histograms (in bar) and density (in line) plots of non-zero expression values from four genes, RGS1, A2M, C3 and FKP5. }
    \label{fig:density}
\end{figure}

Motivated by the astrocytoma scRNA-seq data previously described and also for the purpose of demonstration, we first focus our discussion on the Gaussian distribution $\mathcal{G}(x|
\mu_{ij}, \sigma_j^2) = \exp\left\{-\frac{1}{2}\log2\pi\sigma_j^2-\frac{(X_{ij}-\mu_{ij})^2}{2\sigma_j^2}\right\}$. 
This data set exhibits a high prevalence of dropout events and 75.3\% of expression values are zeros. Figure~\ref{fig:density}a displays the histogram of the standardized non-zero expression values from five hundred randomly selected genes, revealing that its distribution appears symmetric, bell-shaped, and close to a standard normal distribution. Moreover, Figure~\ref{fig:density}b depicts histograms and density plots of the non-zero expression values for four randomly selected genes, RGS1, A2M, C3 and FKBP5. We observe multiple peaks in the distribution of each gene, indicating the existence of cell subgroup structures. 

As shown in a t-SNE plot (Figure~\ref{fig:real-combined}b) of the glioma data, cells from the same sample are mostly clustered together, illustrating the difficulty of identifying biologically meaningful cell subgroups without adjusting for the sample effect. To perform simultaneous cell clustering and differential gene expression detection while accounting for the sample effect (high within-sample correlation), we impose the following structure on the mean expression parameter $\mu_{ij}$,
\begin{eqnarray}
\mu_{ij} = e_{j,k(i)} + \sum_{p=1}^P G_{ip}m_{jp}, 
\label{assmp1}
\end{eqnarray}
Herein, $e_{j,k(i)}$ corresponds to the sample effect for the $j$th gene in sample $k(i)$, where $k(\cdot)$ is the function that gives the sample ID for each cell. For example, $k(i) = k$ indicates that cell $i$ comes from sample $k$; correspondingly, $e_{jk}$ represents the effect of sample $k$ for gene $j$. $G_{ip}$ is the indicator of cell subgroup membership for cell $i$, which is of primary interest to estimate. Specifically, $G_{ip} = 1$ indicates that cell $i$ belongs to hidden subgroup $p$, and $G_{ip'}=0$ for all $p'\ne p$. In addition, $m_{jp}$ represents the corresponding effect of gene $j$ in cell subgroup $p$. For model identification, we set $e_{jK} = 0$ for gene $j = 1,...,J$ in sample $K$. 

Let us denote $Z_{ij} = \operatorname{I}(X_{ij} \ne 0)$. 
The log-likelihood can be written as 
\begin{eqnarray}
\text{log-L} & = & \sum_{j=1}^J \sum_{i=1}^{N} (1-Z_{ij})\log(1-\pi_j) + Z_{ij}\left[\log\pi_j - \frac{1}{2}\left\{\log\left(2\pi\sigma_j^2\right) +  \frac{(X_{ij}-\mu_{ij})^2}{\sigma_j^2}\right\}\right].
\label{log-likelihood}
\end{eqnarray}
The other important task is to identify which genes are the main contributors to cell subgroup differentiation. To achieve this, we further regularize the estimation of $m_{j1}, \cdots,m_{jP}$ via the sparsity-induced penalty on the difference between $m_{jp}$ and $m_{jp'}$, denoted by $\Omega(\bm{m})$. That is, we look to force the trivial genes that are not differentially expressed among the cell subgroups, which are also estimated, to satisfy $m_{j1}=,...,=m_{jP}$. Naturally, the genes which have the largest differences among $m_{jp}$s are likely the main drivers of cell subgroup formation. The penalized negative log-likelihood can be written as, 
\begin{eqnarray}
\ell(X,\boldsymbol{\theta}) \equiv -\text{log-L} + \Omega(\bm{m}) = -\text{log-L} + \frac{\lambda}{P(P-1)}\sum_{j=1}^J\sum_{1\le p<p'\le P} |m_{jp} - m_{jp'}|,  
\label{pl}
\end{eqnarray}
where $\boldsymbol{\theta} = \{\boldsymbol{\pi}, \bm{m}, \bm{e}, \bm{\sigma^2}, \bm{G}\}$ includes all unknown parameters to be estimated, and the tuning parameter $\lambda$ corresponds to the weight of the penalty term $\Omega(\bm{m})$. 

Incorporating the above-described parameter regularization, the new model framework is called Regularized Zero-inflated Mixture Models for scRNA data (RZiMM-scRNA).


\subsection{Parameter Estimation.}
We estimate $\boldsymbol{\theta}$ via minimizing $\ell(X, \boldsymbol{\theta})$, such that
\begin{eqnarray}
\widehat{\boldsymbol{\theta}} = \argmin_{\boldsymbol{\theta}} \ell(X, \boldsymbol{\theta}).
\label{loss}
\end{eqnarray}
First, we obtain the estimate of $\pi_j$, the probability of non-zero values for gene $j$, by solving $\partial \ell(X,\boldsymbol{\theta})/\partial \pi_j =
 0$,
\begin{eqnarray}
\widehat\pi_j = \frac{\sum_{i=1}^n Z_{ij}}{n}. 
\label{eqn:pi}
\end{eqnarray}
Note that minimization of the loss function (\ref{pl}) is convex when either $\bm{G}$ or $\{\bm{m}, \bm{e}, \bm{\sigma^2}\}$ is fixed, but not jointly convex in $\boldsymbol{\theta}$. Therefore, we iteratively estimate $\bm{G}$ and $\{\bm{m}, \bm{e}, \bm{\sigma^2}\}$ conditional on each other, 
by minimizing the following terms separated from $\bm{\pi}$, 
\begin{eqnarray}
\ell(Y,\boldsymbol{\theta}) = l_0(\boldsymbol{\pi}) + \sum_{j=1}^J \sum_{i=1}^N \frac{Z_{ij}}{2}\left\{\log(2\pi\sigma_j^2) + \frac{(X_{ij}-\mu_{ij})^2}{\sigma_j^2}\right\} + \Omega(\bm{m}) = l_0(\boldsymbol{\pi}) + l_1 + \Omega(\bm{m}).  
\end{eqnarray}
$l_1$ can be written as, 
\begin{eqnarray}
l_1 = \sum_{j=1}^J \frac{1}{2}\left\{T_j\log(2\pi\sigma_j^2) + \frac{\left(X_j' - G^{(j)}\bm{m} - S^{(j)}\bm{e}\right)^T\left(X_j' - G^{(j)}\bm{m} - S^{(j)}\bm{e}\right)}{\sigma_j^2}\right\},
\end{eqnarray}
Herein, $T_j = \sum_{i=1}^n Z_{ij}$, $X_j'$ is a vector of $X_{ij}$'s when $Z_{ij} = 1$, and $G$ is a $n\times P$ matrix for the group allocation. The $S$ is a $n\times (K-1)$ matrix, with the element $S_{ik} = 1$ when cell $i$ belongs to sample $k$, and 0 otherwise. The $G^{(j)}$ and $S^{(j)}$ include only rows corresponding to $Z_{ij} = 1$.

Since the penalty term $\Omega(\bm{m})$ is non-differentiable, we adopt Majorization Maximization (MM) algorithm \citep{lange2013mm} in the optimization procedure, 
\begin{eqnarray}
\Omega(\bm{m}) \le \frac{\lambda}{P(P-1)}\sum_{j=1}^J\sum_{1\le p < p'\le P} \frac{1}{2}\left\{\frac{(m_{jp} - m_{jp'})^2}{b_{jpp'}} + b_{jpp'}\right\} = \Omega'(\bm{m}, \bm{b})
\end{eqnarray}
where $b_{jpp'} > 0$, $\Omega'(\bm{m}, \bm{b})$ is differentiable and the equivalence holds when $b_{jpp'} = |m_{jp} - m_{jp'}|$. 

Conditional on $G$, we can obtain the estimate of $\bm{m}$ and $\bm{e}$ iteratively as follows, 
\begin{enumerate}
    \item Set $b_{jpp'} = |\widehat{m}_{jp} - \widehat{m}_{jp'}| + \delta$, where $\delta > 0$ to avoid division by zero. 
    \item Estimate $\widehat{\bm{m}}$ and $\widehat{\bm{e}}$ by setting $\frac{\partial \left\{(I) + \Omega'(\bm{m}, \bm{b})\right\}}{\partial (\bm{m}, \bm{e})} \equiv 0$. 
\end{enumerate}
We follow this by solving $\frac{\partial (I)}{\partial \bm\sigma^2} \equiv 0$ to estimate $\widehat{\bm\sigma^2}$. 
Then, conditional on the estimate of $\bm{m}$, $\bm{e}$ and $\bm\sigma^2$, for $i = 1,...,n$, we estimate $G_i$ in the following, 
\begin{eqnarray}
\widehat{p}(i) = \argmin_{p} \sum_{j=1}^J \frac{\left(X_{ij}-\widehat{m}_{jp}-\widehat{e}_{j,k(i)}\right)^2}{2\widehat\sigma_j^2},
\label{eqn:gi}
\end{eqnarray}
where $\widehat{G}_i = (\widehat{G}_{i1}, ..., \widehat{G}_{iP})$, $\widehat{G}_{ip} = 1$ when $p=\widehat{p}(i)$, and 0 otherwise. 

The complete procedure can be summarized as follows, \begin{enumerate}
    \item Estimate $\widehat{\bm\pi}$ via (\ref{eqn:pi}). Initialize $\widehat{G}$ by randomly assigning cell $i$ to group $p$, $1\le p \le P$. 
    \item Iterate the following procedure until convergence or the maximum number of iterations is reached. 
    \begin{enumerate}
        \item Fix $\widehat{G}$, and estimate $\widehat{\bm{m}}$, $\widehat{\bm{e}}$ and $\widehat{\bm\sigma}^2$ via MM algorithm. 
        \item Fix $\widehat{\bm{m}}$, and estimate $\widehat{\bm{e}}$ and $\widehat{\bm\sigma}^2$, and $\widehat{G}$ via (\ref{eqn:gi}). 
    \end{enumerate}
\end{enumerate}

The tuning parameter $\lambda$ of the penalty term $\Omega(\bm{m})$ and the optimal number of cell types can be automatically determined based on the modified BIC proposed by \citeauthor{wang2009shrinkage}, 
$$
\operatorname{BIC}_\gamma = -2\operatorname{log-L} + \gamma d \log N$$
where $d$ is the number of total parameters, $\gamma > 1$ is a pre-specified parameter. We set $\gamma = \log(\log J)$ as suggested by \citeauthor{wang2009shrinkage}. As shown by \citeauthor{wang2009shrinkage}, this modification improves the performance over the original BIC, which tends to obtain over-parsimonious results on clustering.

\subsection{Gene Importance Scoring. }
The new model framework not only achieves cell subgroup (type) discovery in a data-adaptive fashion, but also provides meaningful insight in detection of important genes.  To identify the differentially expressed genes which drive  formation of different cell types by RZiMM-scRNA, we define an importance score for each gene as the summation of pairwise differences between cell types, 
\begin{eqnarray}
\operatorname{IS}_j = \frac{1}{P(P-1)}\sum_{1\le p<p' \le P}|m_{jp} - m_{jp'}| 
\end{eqnarray}
Genes with large values of $\text{IS}_j$ are considered to be cluster-distinctive biomarker candidates.

\subsection{Cell-Type-Specific Driving Differential Gene Expression. }
It is also straightforward to utilize RZiMM-scRNA to identify genes that are non-trivially over-expressed or under-expressed in a specific cell type. We define the measure of differential expression level for gene $j$ in cell type $p$ as,
\begin{eqnarray}
\operatorname{DE}_{jp} = \frac{1}{P-1}\sum_{p'\ne p}\left(m_{jp} - m_{jp'}\right) 
\label{diff}
\end{eqnarray}
Genes with the largest magnitudes of $\text{DE}_{jp}$ have the most over-expression or under-expression in cell type $p$, and are viewed as the main contributors to separate cell type $p$ from the others. This measure is useful in aligning cell clusters to known cell types and identifying new biomarkers of a specific cell type. 

\subsection{A remark} The focus of this paper is to introduce and apply a new joint-sample modeling framework for simultaneous discovery of cell subtypes and detection of driving gene markers. This new system yields an intuitive measure of gene importance, which can be used to directly rank genes according to their strength of association with the discovered cell subgroups. Statistical significance can also be assessed through bootstrapping, which will be deferred to our future investigation.    


\section{Results}

We conduct extensive simulation studies to assess the performance of RZiMM-scRNA in some common settings and conclude with an application to the previously described astrocytoma scRNA-seq datasets. 

In addition to RZiMM-scRNA, we also implement Hierarchical Clustering (H-Clust), K-means, Seurat, SC3, fastMNN, and RZiMM not adjusting for within-sample effect (RZiMM-Naive) for comparison. H-Clust and K-Means are used to cluster the cells and ANOVA are then conducted using the ``anova'' function in R package ``stats'' \citep{statsR} to detect significant genes based on likelihood-ratio test. FastMNN extracts fifty top features in the PC space based on PCA with batch effects corrected and then performs cell clustering via K-means. The gene statistical significance level is then determined in the same way as in H-Clust and K-Means. Seurat and SC3 have their own pipelines to identify differentially expressed genes. 

K-means and H-Clust are implemented using the functions ``kmeans'' and ``hclust'', respectively, in R package ``stats''. For H-Clust, we use one minus Pearson correlation as the dissimilarity measurement coupled with average agglomeration. We then use ``cutree'' function in R package ``stats'' to divide the ordered cells into clusters.  For K-Means, Euclidean distance is used as the dissimilarity measurement. FastMNN is implemented using function ``fastMNN'' in R package ``batchelor''. Seurat and SC3 are implemented using R packages ``Seurat'' and ``SC3''.

\subsection{Simulation Study}
\subsubsection{Expression Data from Zero-inflated Gaussian Distribution}
In this study, we generate zero-inflated expression data from models (\ref{assump0}) and (\ref{assmp1}), assuming $\mathcal{G}$ is a Gaussian distribution, which is motivated by the first brain cancer study.  We consider $N = 1000$ cells, $J = 1000$ genes, $K = 10$ samples, and $P = 4$ hidden cell subgroups, where sample ID $k(i)$ and subgroup membership $p(i)$ are randomly assigned. Each of the genes $1\le j\le 100$ is a differentially expressed (DE) gene, where a randomly selected cell subgroup 
is designed to have a higher mean expression than the other three subgroups by a value drawn from the uniform distribution $\U(\delta, 2\delta)$, while the other three groups have a constant effect $m_j$ drawn from $\U(4, 8)$. The effect size, $\delta$ ranging from 0.5 to 1.5, indicates the difference in DE genes between subgroups. For genes $100 < j \le J$, we set 
$m_{j1}=...=m_{jP} = m_j$ where $m_j \sim \U(4, 8)$. That is, these genes are not differentially expressed among the cell subgroups.  In each scenario of $\delta$ value, one hundred simulated data sets are used for investigation. The gene-specific sample effect $e_{jk}$ is drawn from the normal distribution $\N(0, 0.2)$, for $k=1,...,K$, $j=1,...,J$. Therefore, the mean expression $\mu_{ij} = m_{j, p(i)} + e_{j, k(i)}$. Further, the probability of zero-expression $\pi_{ij} = \operatorname{expit}(\mu_{ij}/2-4)$, and the gene expression intensity values are generated as $X_{ij}=Z_{ij}W_{ij}$, where $Z_{ij}\sim\operatorname{Bin}(1,\pi_{ij})$ and $W_{ij}\sim\operatorname{N}(\mu_{ij}, 1)$. 
The within-sample correlation structure at $\delta = 0.5$ is visualized in Supplement Figure S1, where the positive correlation among cells within a sample (block-wise) is clearly observed.



\begin{figure}
    \centering
    \includegraphics[width = .7\textwidth]{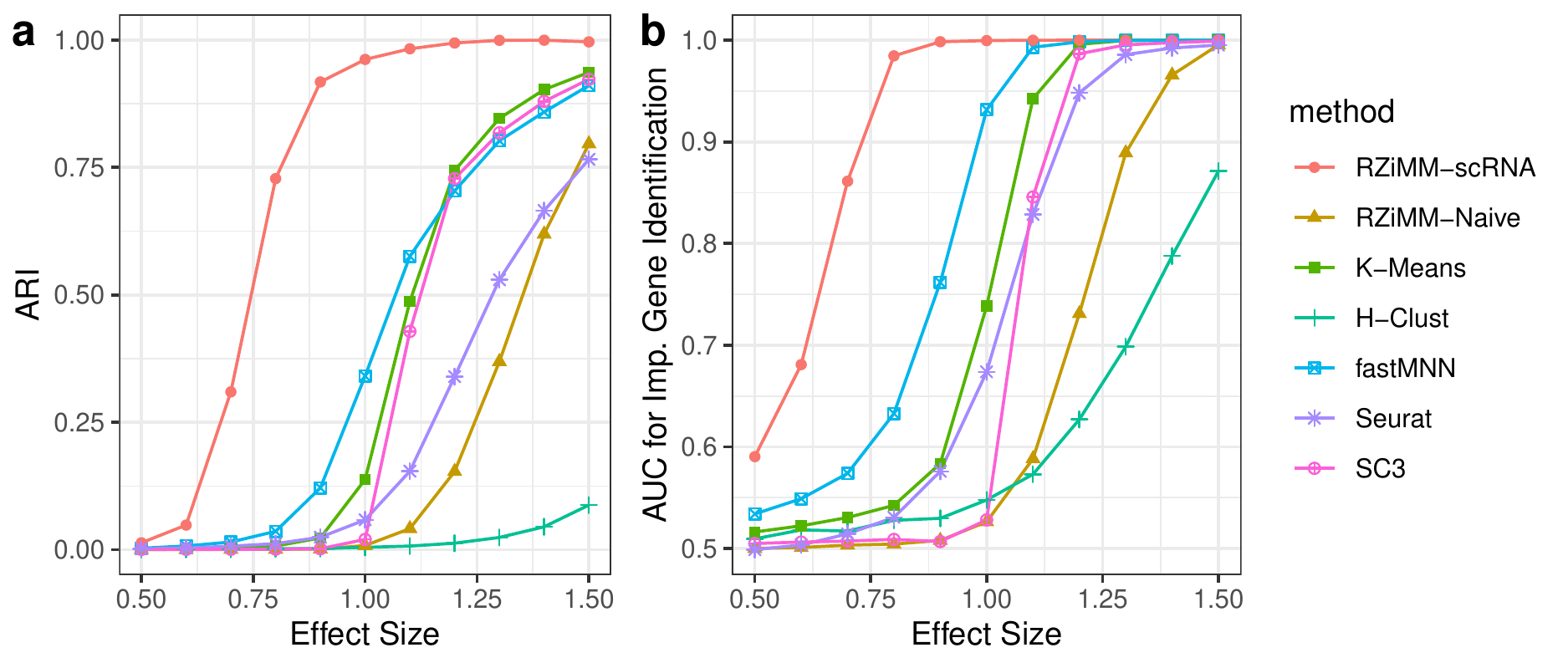}
    \caption{
    Performance Plots for the Simulation Study with Normalized Data. a. ARI for the clustering result. b. AUC for important gene identification.  }
    \label{fig:sim_perf_comb}
\end{figure}

We assume four cell clusters (also the truth) for all methods, as our focus is to compare cell subgroup composition and biomarker detection among the various methods. 
To evaluate the clustering performance, we use the Adjusted Rand Index (ARI) \citep{hubert1985comparing}.
In addition, we use the area under ROC curve (AUC) as the measure of biomarker detection accuracy, where larger values of AUC correspond to higher biomarker detection accuracy. Figure~\ref{fig:sim_perf_comb}a-b plots the clustering and biomarker detection accuracies for each of the methods averaged over the one hundred simulated data sets against various values of $\delta$. It is clear that RZiMM-scRNA consistently outperforms all the other methods in both clustering and biomarker detection accuracy. 
As the effect size $\delta$ increases, the ARI for clustering and AUC for DE gene identification from all methods are improved. The value of $\delta$ needed for RZiMM-scRNA to reach perfect ARI and AUC is around 1, and that for other methods is greater than 1.5. This promising performance on zero-inflated expression data drawn from the Gaussian distribution suggests that RZiMM-scRNA is a powerful method to analyze the astrocytoma scRNA-seq expression data. 

\subsubsection{Count Data from Zero-inflated Poisson Distribution}
In this study, the popular Poisson distribution is considered to mimic the count nature of the data generated by single-cell RNA-sequencing. Similar to the previous simulation setting, we consider $N = 1000$ cells, $J = 1000$ genes, $K = 10$ samples, and $P = 4$ hidden cell subgroups. Each of the genes $1\le j\le 100$ is a differentially expressed (DE) gene, where a randomly selected cell subgroup $p^{(j)}$ is designed to have a higher mean expression than the other three subgroups by a value drawn from the uniform distribution $\U(\delta, 2\delta)$, while the other three groups have a constant effect parameter $m_j$ drawn from $\U(4, 8)$. The effect size, $\delta$ ranging from 1 to 5, indicates the degree of difference in DE genes between subgroups. For genes $100 < j \le J$, we set 
$m_{j1}=...=m_{jP} = m_j$ where $m_j \sim \U(4, 8)$. The gene-specific sample effect $e_{jk}$ is allowed to vary, drawn from the normal distribution $\N(0, 2)$, for $k=1,...,K$, $j=1,...,J$. The mean expression $\mu_{ij} = \max(m_{j, p(i)} + e_{j, k(i)}, 0.5)$. Further, the probability of zero-expression $\pi_{ij} = \operatorname{expit}(\mu_{ij}/6-2)$, and thus $X_{ij}=Z_{ij}W_{ij}$, where $Z_{ij}\sim\operatorname{Bin}(1,\pi_{ij})$ and $W_{ij}\sim\operatorname{Poisson}(\mu_{ij})$, resulting in the count data following zero-inflated Poisson distributions.

As discussed in Section 3, the proposed model framework can flexibly accommodate non-Gaussian distributed data. We use RZiMM-Pois as the abbreviation of out method implemented under zero-inflated Poisson distribution. We are also interested to examine the performance of adopting zero-inflated negative binomial distribution (RZiMM-NB) and zero-inflated Gaussian distribution (RZiMM-scRNA
). From the comparison, we hope to assess the robustness of results to various distributions. Note that the result under the true Poisson distribution (RZIMM-Pois) is treated as the benchmark.  

\begin{figure}
    \centering
    \includegraphics[width = 0.8\textwidth]{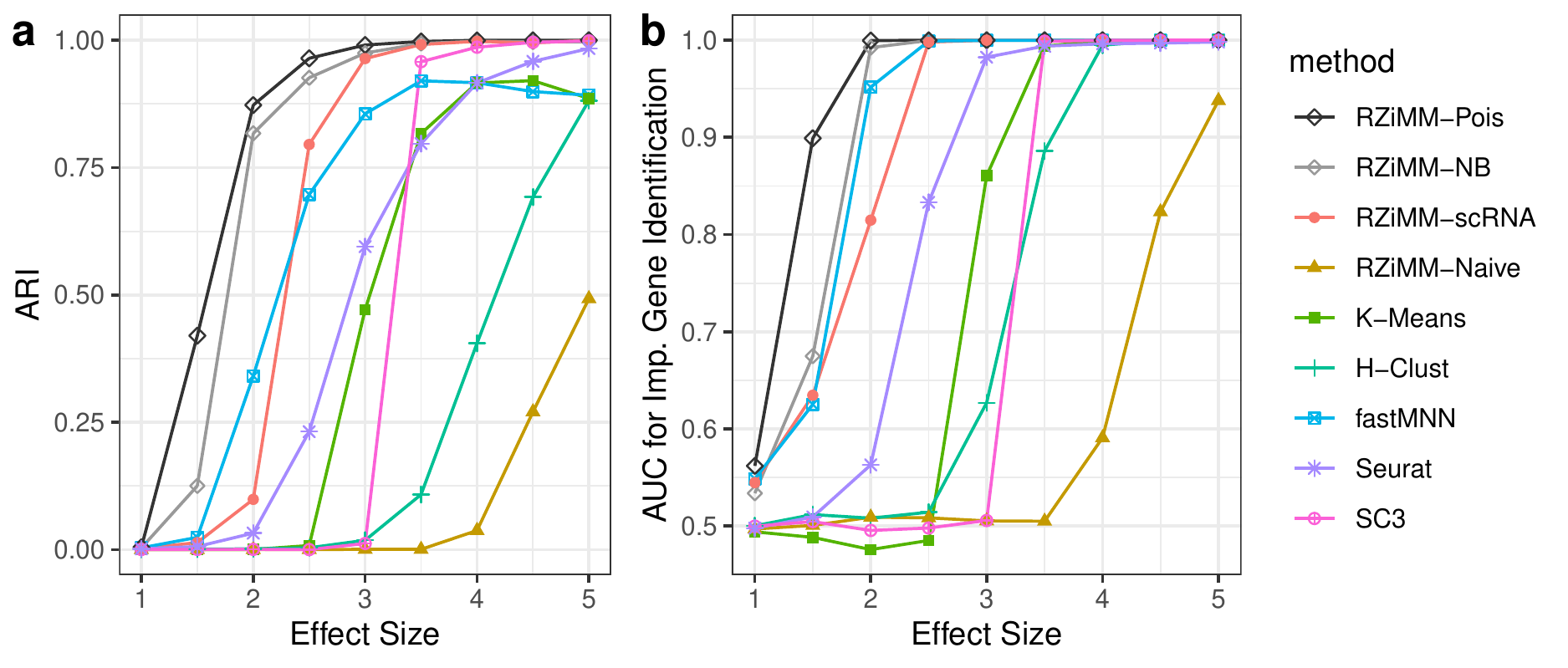}
    \caption{Performance Plots for the Simulation Study with Count Data. a. ARI for the clustering result. b. AUC for important gene identification. }
    \label{fig:sim-count}
\end{figure}

Figure \ref{fig:sim-count} displays the ARI and AUC results of all the considered methods. Note that RZIMM-NB's performance is very close to the benchmark RZIMM-Pois, which is not surprising due to their intrinsic mathematical connection. Interestingly, the next best performers are RZIMM-scRNA and fastMNN. The fairly good performance of RZIMM-scRNA suggests that the Gaussian distribution works robustly and reasonably well, regardless of the true underlying data distributions. In addition, both popular pipeline methods Seurat and SC3 demonstrate inferior performance. Overall, the result demonstrates that the RZiMM model framework is a powerful substitute to the most common alternatives of scRNA-seq data analysis.

\subsection{A Semi-simulation Scenario based on the Real Astrocytoma Data}
We would like to make the simulated data as close to the reality as possible, while being able to generate some truth structure for performance assessment. Instead of generating the data from parametric models, we fully utilize the first astrocytoma scRNA-seq dataset via permutation. Specifically,
we first randomly permute expression intensities of each gene across cells within a sample, as to remove the unknown underlying  cell subgroup structure while preserving the within-sample correlation. Next, we add the true structure of cell subgroups and the corresponding driving gene biomarkers as follows. We randomly divide the cells into four clusters and select one hundred genes as the pseudo-important biomarkers. For each of these important genes, we elevate the expression intensities in a randomly assigned cell subgroup by a positive value drawn from the uniform distribution $\operatorname{U}(v, v+1)$, where several values of $v$ between $0$ and $4$ are examined. 

Note that $v$ corresponds to the effect size of the hidden cell subgroups, where larger values of $v$ correspond to more distinct cell subgroups and less difficulty for a method to correctly identify the underlying cell subgroup structure. The cell correlation map at $v = 4$ is presented in Supplement Figure S2, in which cells are ordered by sample and then by true hidden subgroup within each sample. The within-sample correlation is quite noticeable in the glioma data, as shown in different blocks. For example, the second sample to the left MGH61 (shown in gray) reveals a strong between-cell correlation. The small blocks in the map manifest the non-trivial correlation within a cell subgroup. 

We again compare the clustering and biomarker detection accuracies across methods (see Figure~\ref{fig:sim_perf_comb}c-d). We observe that RZiMM-scRNA is the most powerful of all the methods, as it starts accurately detecting the underlying true cell subgroup structure with the smallest effect size $v$. Specifically, we observe that RZiMM-scRNA can detect true cell clusters with 100\% accuracy for effect sizes of at least $1.1$. In contrast, 
Seurat can only achieve perfect accuracy when the effect size is greater than $4$. Since SC3, RZiMM-Naive, K-Means, and H-Clust are unable to account for the sample effect, they completely fail to detect the true cell subgroups when $v\le 3$ and cannot identify the important genes (low AUC in the right panel). Only as the effect size becomes very large (specifically, $v = 4$) can RZiMM-Naive fully detect the true subgroup structure and important genes, whereas SC3, K-Means and H-Clust cannot improve their performance. 

This study also suggests that RZiMM-scRNA, compared to the other methods, will likely provide more meaningful insight from the astrocytoma data, as the empirical distributions of gene expression data in that real study are essentially preserved in our simulation. 

\begin{figure}
    \centering
    \includegraphics[width = 0.8\textwidth]{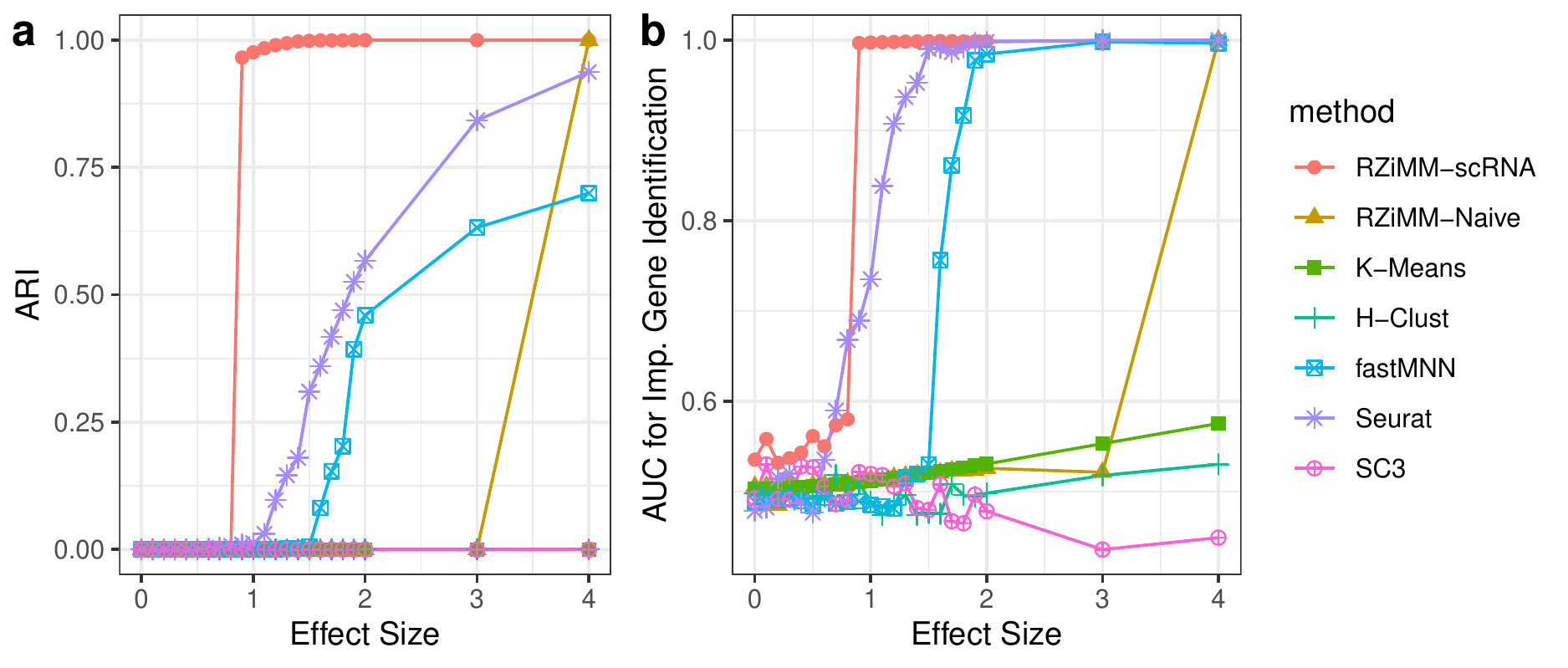}
    \caption{Performance Plots for the Semi-simulation Study. a. ARI for clustering. b. AUC for important gene identification. }
    \label{fig:semi-sim}
\end{figure}



\subsection{Analysis of Astrocytoma Data}

Finally, we apply each of the methods to the first astrocytoma scRNA-seq dataset. In the study done by \citeauthor{venteicher2017decoupling}, ten IDH-mutant human astrocytoma tumors spanning grades II to IV were disaggregated, sorted into single cells, and profiled by Smart-seq2 \citep{picelli2014full}. Expression levels were quantified as $\E_{jk} = \log_{2}(\text{TPM}_{jk}/10+1)$, where $\text{TPM}_{jk}$ refers to  transcript-per-million for gene $j$ in sample $k$, as calculated using the software RSEM (RNA-Seq by Expectation-Maximization, \cite{li2011rsem}). Cells with fewer than 3,000 detected genes or an average  expression level of a curated list of housekeeping genes below 2.5 were excluded. Consequently, 6,341 single cells were retained in the study. In accordance with the analysis conducted by \citeauthor{venteicher2017decoupling}, we include 8,231 genes among the total of 23,686 for further analysis, whose expression levels satisfy $\log_{2}[\text{Average}(\text{TPM}_{j,1...N}) + 1] \geq 4$. 

An important first task we must address is determining the optimal number of cell clusters for each method. For RZiMM-scRNA, we adopt modified BIC \citep{wang2009shrinkage} as the model selection criterion integrated with the model-based method, with a grid search over $P \in \{2, 3, ..., 15\}$ and $\lambda \in \{0, 20, 40, ..., 200\}$ (see Supplement Figure S4). As the result, we obtain $P=6$ as the optimal number of clusters. For K-Means, H-Clust, and fastMNN, we utilize average silhouette value \citep{rousseeuw1987silhouettes} and gap statistics \citep{tibshirani2001estimating}, examining the range of $P$ from $2$ to $15$. The average silhouette value method is implemented using the ``fviz\_nbclust'' function in R package ``factoextra'' \citep{factoextraR} and the gap statistic is implemented using MATLAB package ``clustering.evaluation''. The average silhouette value method suggests $2$ cell clusters for the three methods, which is too small to capture the important pattern in our data. On the other hand, the gap statistics suggests $6$, $15$, and $14$ clusters for K-Means, H-Clust and fastMNN, respectively (see Supplement Figure S5). Taking into account the discordance between the average silhouette value and gap statistic suggestions, and for fair comparison, we use $6$ clusters for these methods.

\begin{figure}
    \centering
    \includegraphics[width = \textwidth]{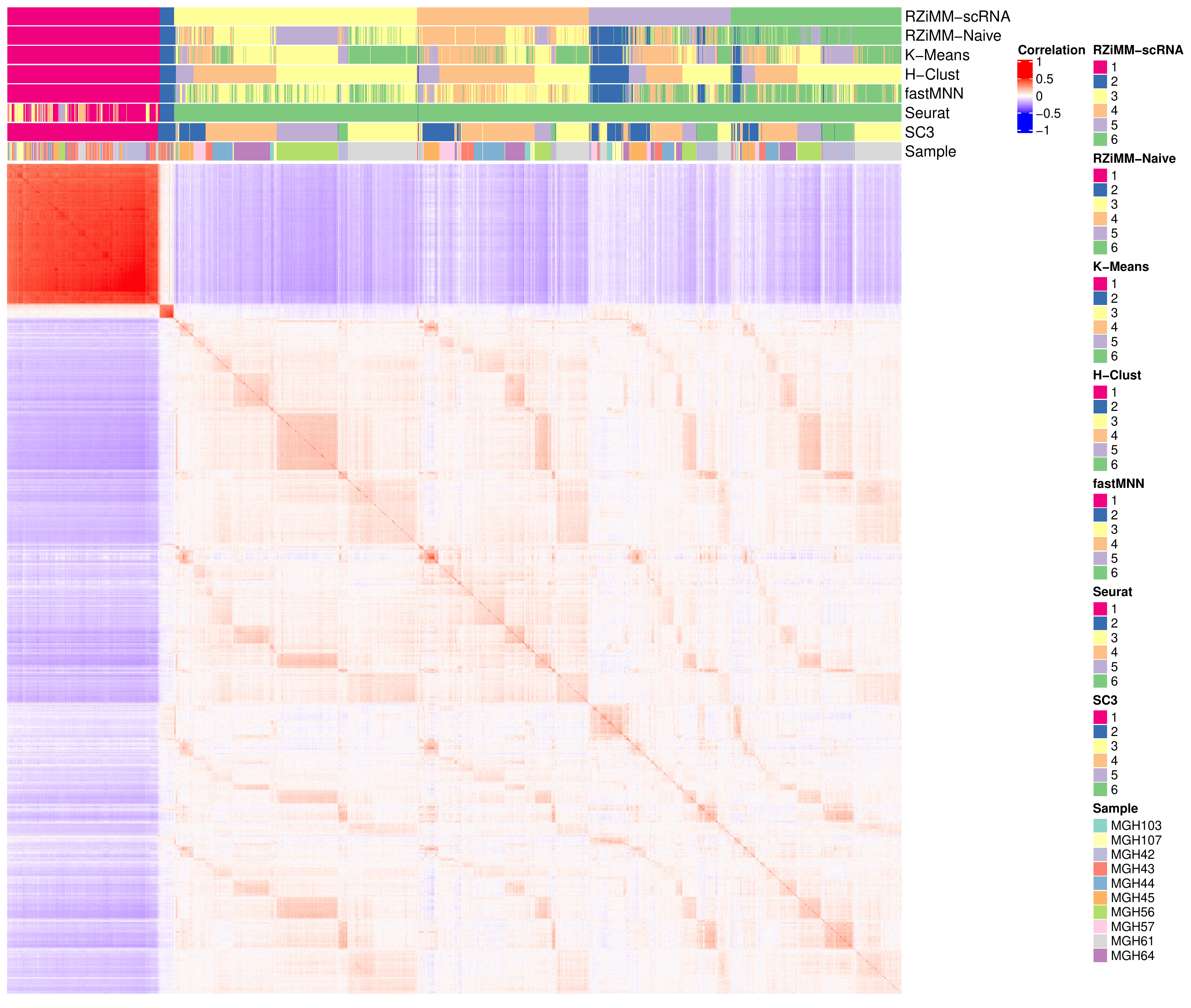}
    \caption{Correlation Map for Astrocytoma scRNA-seq Data. The genes are ordered by RZiMM-scRNA and H-Clust. }
    \label{fig:corr}
\end{figure}

\begin{figure}
    \centering
    \includegraphics[width = .95\textwidth]{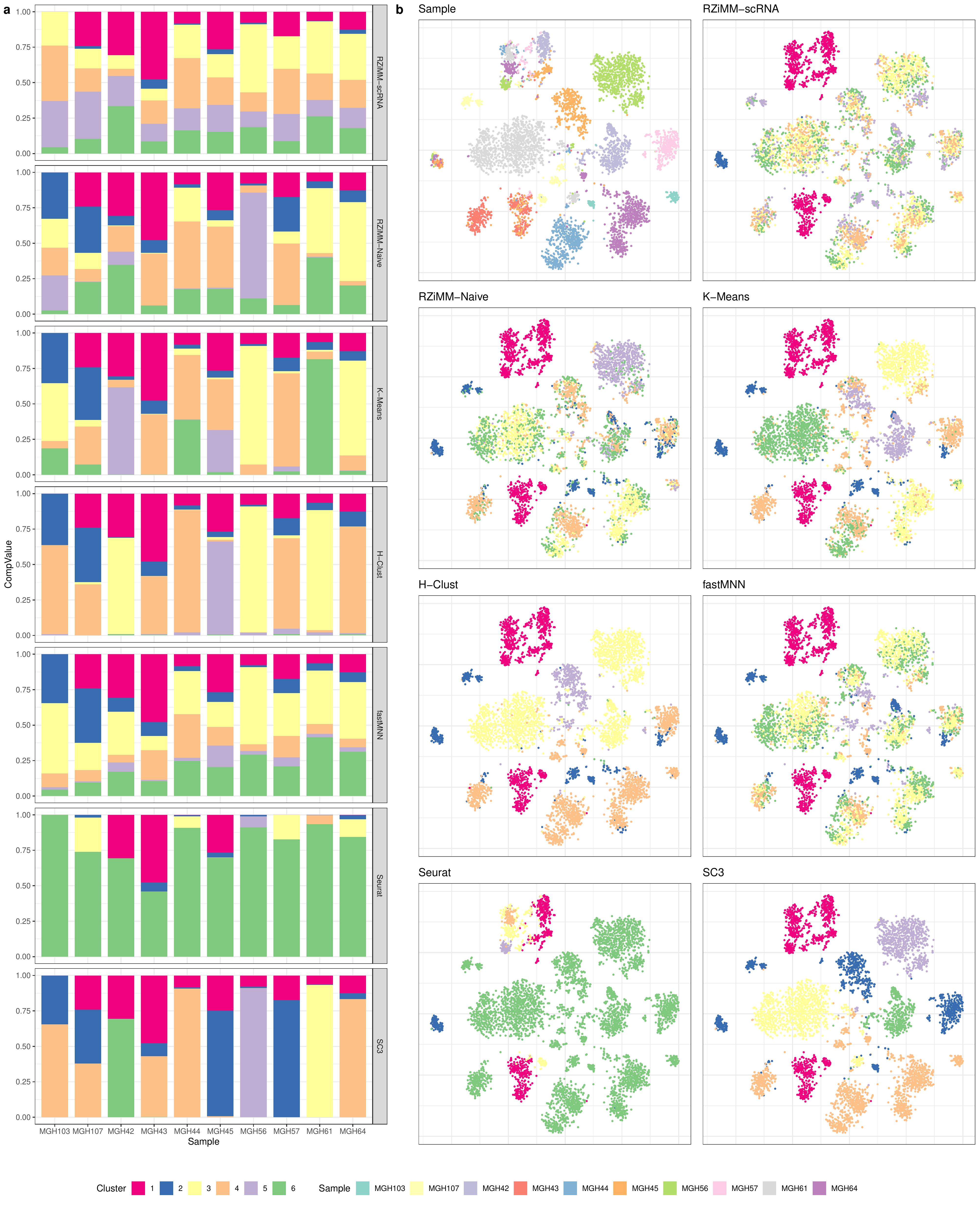}
    \caption{Clustering Analyses of the Astrocytoma Data. a. Compositional plots for cluster membership of each sample estimated by the five methods. b. t-SNE plots colored by sample or clustering methods. }
    \label{fig:real-combined}
\end{figure}

Figure~\ref{fig:real-combined}a is a compositional plot displaying the proportions of cells clustered into each cell group for each sample. We notice that for several patient samples, such as MGH56 and MGH61, SC3, K-Means and H-Clust tend to cluster cells together by sample. This is also observed for RZiMM-Naive and SC3. For example, the cells of sample MGH56 are mostly clustered into Cluster 5 (purple) by RZiMM-Naive and SC3, or Cluster 3 (yellow) by both K-Means and H-Clust. Once again, we see the importance of addressing sample effect. Without the ability to handle the sample effect, the clustering patterns of RZiMM-Naive, SC3 K-Means, and H-Clust are dominated by sample identity, rather than biologically important distinctions. These same clustering patterns are observed in a correlation map (Figure~\ref{fig:corr}), which reveals prominent within-sample correlations, and in t-SNE plots (Figure~\ref{fig:real-combined}b), where cells appear to be clustered by sample identity.


\begin{figure}
    \centering
    \includegraphics[width = .5\textwidth]{xxx-pair-ari.pdf}
    \caption{Pairwise ARI for Sample Index and Clustering Results}
    \label{fig:pair-ari}
\end{figure}

The cell correlation map (Figure~\ref{fig:corr}) presents the pairwise Pearson correlation between cells and the heatmap (Supp Figure~S3) shows the expression intensity of each gene in each cell, both of which are annotated with sample labels and the clustering membership estimated by each method. The cells are first ordered by the six clusters identified by RZiMM-scRNA, and then ordered within each cluster by H-Clust. Note that Cluster 1 from RZiMM-scRNA (red) is identified by six clustering methods, as it contains highly correlated cells, but not by Seurat as shown by the correlation map in Figure~\ref{fig:corr}. The heatmap in Supp Figure~S3 also reveals that a large group of genes are highly expressed in Cluster 1, which is made easily identifiable by all the clustering methods. Interestingly, Cluster 1 matches the microglial group of cells reported in \citeauthor{venteicher2017decoupling}. Microglial cells mediate immune responses in the central nervous system and are re-programmed by tumor cells to create a favorable microenvironment for cancer progression \citep{abels2019glioblastoma}. 
Moreover, Cluster 2 from RZiMM-scRNA (blue) and Seurat is composed of a small group of highly correlated cells, as shown in the correlation map. In fact, Cluster 2 matches another biologically meaningful cell type in the oligodendrocyte group reported by \citeauthor{venteicher2017decoupling}. Oligodendrocytes produce myelin in the central nervous system and its precursor cells help modulate particular microenvironments for glioma cells \citep{hide2018oligodendrocyte}. However, none of the aforementioned methods other than RZiMM-scRNA or Seurat are able to distinguish this cell type from the other cells in our data. Specifically, these other methods tend to combine cells in Cluster 2 with some cells in Clusters 5 and 6, as identified by RZiMM-scRNA, into the same subgroup, indicated by blue for all the other methods.

We also visualize this result using two-dimensional t-SNE plots in Figure~\ref{fig:real-combined}b. In the top left panel, the plot is colored by sample; while in other panels, the plot is colored by the clustering result for each method. Again, Cluster 2 is revealed to be clearly separated from the other cells. However, only RZiMM-scRNA and Seurat can sharply identify this pattern. All the other methods pool many other cells, which appear to be distant, into this cluster (blue). This demonstrates the superior performance of RZiMM-scRNA, as well as Seurat in identifying rare cell types. 

In Figure~\ref{fig:pair-ari}, pairwise ARIs across different clustering methods and sample ID are presented. It is noteworthy that when comparing RZiMM-scRNA, fastMNN and Seurat results to sample ID, the ARI is close to zero, while other methods have fairly high ARIs to sample ID. This result indicates that RZiMM-scRNA, fastMNN and Seurat can effectively remove the batch effects, while the other methods fail. In addition, RZiMM-scRNA shows better agreement with RZiMM-Naive and fastMNN, indicated by fairly large ARI values. In contrast, Seurat has overall low agreement with all the other methods.

\begin{figure}
    \centering
    \includegraphics[width = \textwidth]{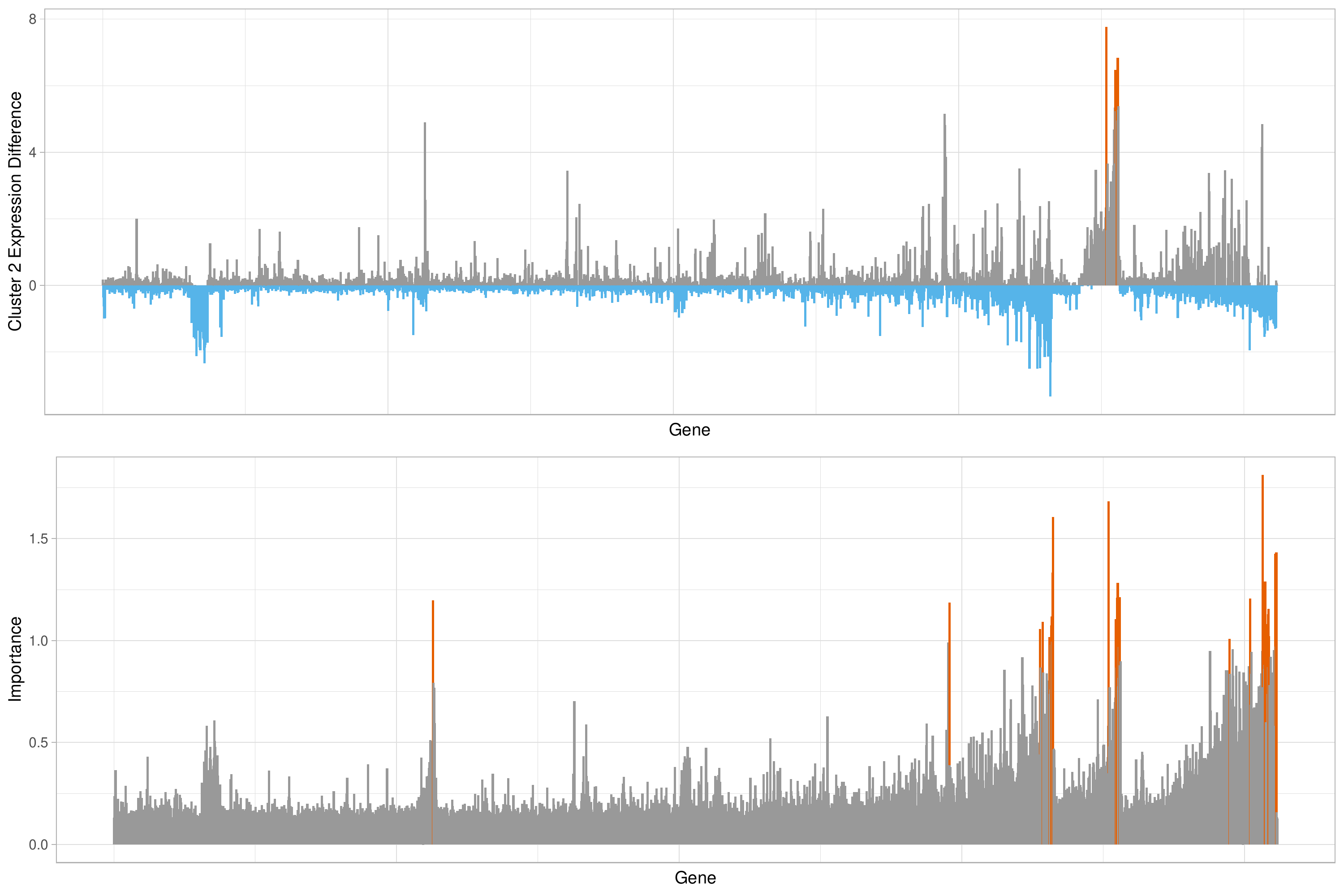}
    \caption{RZiMM-scRNA Importance. Top panel: mean difference in normalized expression between Cluster 2 and the rest (orange: top five over-expressed genes in Cluster 2, PLP1, CLDN11, MOG, MBP and TF; grey: over-expressed genes except the top five; blue: the under-expressed genes in Cluster 2). Bottom panel: importance scores for differentially expressed genes (orange: top forty differentially expressed genes; grey: other genes)}.
    \label{fig:imp}
\end{figure}

For the particularly interesting Cluster 2 corresponding to the oligodendrocyte cell type, we also compute the measure of differential gene expression between Cluster 2 and all the other clusters according to equation (13) in RZiMM-scRNA, which is displayed in the top panel of Figure~\ref{fig:imp}. The under-expressed genes in Cluster 2 are colored in blue, the top five over-expressed genes in Cluster 2 ($\operatorname{DE}_{j2} > 6$) are colored in orange, and the rest are colored in gray. We observe a set of genes highly over-expressed in Cluster 2 with the differential expression measure over four. In particular, the top five over-expressed genes in Cluster 2, PLP1, CLDN11, MOG, MBP and TF, are colored in orange. These five genes are also among the top forty important genes identified by RZiMM-scRNA, which yield the overall largest differentiation across the six cell clusters, presented in the bottom panel of Figure~\ref{fig:imp} and Supplement Table S1. In Supplement Table S1, we observe that these forty genes are either over-expressed in Cluster 1 or 2 or under-expressed in Cluster 1 or 2, as compared to all the other clusters. 



Furthermore, we investigate the biological significance of the identified top forty genes. Promisingly, proteins encoded by most of these genes play roles in the nervous system ranging from immune response to cell migration and proliferation, and many are associated with forms of brain cancer, brain disease, and other ailments. We include a brief description of the biological significance of some of these genes in Table~\ref{tab:imp}, specifically, their relationship to cancers such as low-grade glioma (LGG), high-grade glioma (HGG), and glioblastoma (GBM), and neurological and neurodegenerative diseases. For a more extensive account of the function and biological significance of these genes, please refer to Table S2 in the Supplement.

\footnotesize
\begin{center}
\setlength\arrayrulewidth{0.01pt}
\begin{longtable}{|c|c|c|}
\caption{Top Important Genes}\label{tab:imp}\\
\hline
\textbf{Gene} & \textbf{Significance} & \textbf{Ref.} \\
\hline
\endfirsthead
\multicolumn{3}{c}%
{\tablename\ \thetable\ -- \textit{Continued from previous page}} \\
\hline
\textbf{Gene} & \textbf{Significance} & \textbf{Ref.} \\
\hline
\endhead
\hline \multicolumn{3}{r}{\textit{Continued on next page}} \\
\endfoot
\hline
\endlastfoot
SPP1 & \makecell{- Upregulated by glioma-associated \\ microglia/macrophages (GAMs) and associated \\ with poor patient prognosis in LGG and GBM} &
\makecell{\cite{chen2019identification} \\ \cite{szulzewsky2015glioma}} \\
\hline
PLP1 & \makecell{- Key for glioma development \\ and associated with patient prognosis} & \makecell{\cite{chen2019screening}} \\
\hline
BCAN & \makecell{- May promote the growth and \\ motility of brain tumor cells} & 
\makecell{\cite{o2016reference}} \\
\hline
RGS1 & \makecell{- Gene polymorphisms associated with MS} & 
\makecell{\cite{caballero2016autoimmunity} \\ \cite{sawcer2011genetic}} \\
\hline
CD74 & \makecell{- Upregulated in HGG and \\ positively associated with patient survival} & \makecell{\cite{zeiner2015mif}} \\
\hline
GPM6A & \makecell{- Identified as a top gene \\ candidate as a causal gene for SZ} & \makecell{\cite{ma2018integrated}} \\
\hline
PTPRZ1 & \makecell{- Upregulated in LGG and GBM, and associated with \\ malignant growth of GBM and poor patient prognosis} & \makecell{\cite{xia2019expression} \\ \cite{shi2017tumour}} \\
\hline
CCL3 & \makecell{- Downregulated in HGG} & \makecell{ \cite{vidyarthi2019predominance}} \\
\hline
SEPP1 & \makecell{- Inactivation causes severe neurological dysfunction \\ and brainstem neurodegeneration in mice} & \makecell{\cite{byrns2014mice}} \\
\hline
GSN & \makecell{- Significantly downregulated in \\ many cancers including GBM} & \makecell{\cite{miyauchi2018identification}} \\
\hline
FXYD6 & \makecell{- Downregulated in brain of AD mouse model} & \makecell{\cite{george2010serial}} \\
\hline
MOG & \makecell{- Anti-MOG antibodies associated with demyelinating \\ diseases such as acute disseminated encephalomyelitis \\ and neuromyelitis optica spectrum disorder} & \makecell{ \cite{sinmaz2016mapping}} \\
\hline
MBP & \makecell{- Increased expression levels detected in AD cortex \\ and could be involved in amyloid plaque formation} & \makecell{\cite{zhan2015myelin}} \\
\hline
CCL4 & \makecell{- Upregulated by GAMs} & \makecell{\cite{zeiner2019distribution}} \\
\hline
A2M & \makecell{- Inhibits proliferation, migration,\\ and invasion of astrocytoma cells} & \makecell{\cite{lindner2010alpha2}} \\
\hline
APOE & \makecell{- Upregulated in grade II-IV astrocytoma} & 
\makecell{\cite{rousseau2006expression}\\} \\
\hline
CRYAB & \makecell{- Upregulated in IDH1$^\text{R132H}$ mutant \\ anaplastic astrocytoma and GBM} & \makecell{
\cite{avliyakulov2014c} \\ \cite{kore2014inflammatory}} \\
\hline
CLDN11 & \makecell{- Absence in CNS myelin perturbs behavior in \\ mice and could lead to neuropsychiatric disease} & \makecell{\cite{maheras2018absence}} \\
\hline
FEZ1 & \makecell{- Abnormal, unphosphorylated co-aggregates \\ present in brains of transgenic AD mice models} & \makecell{\cite{butkevich2016phosphorylation}} \\
\hline
HLA-E & \makecell{- Upregulated in grade II-IV astrocytoma} & \makecell{ \cite{mittelbronn2007elevated}} \\
\hline
TF & \makecell{- May play role in increased \\ iron deposition in MS brain} & \makecell{\cite{khalil2014cerebrospinal}\\} \\
\hline
HLA-DRA & \makecell{- Low expression levels associated with poor \\ prognosis in pediatric adrenocortical tumors} & \makecell{\cite{leite2014low}} \\
\hline
C1orf61 & \makecell{- May participate in occurrence \\ and development of GBM} & \makecell{\cite{jia2019integrative}} \\
\hline
APLP1 & \makecell{- May contribute to pathogenesis of \\ AD-associated neurodegeneration} & \makecell{
\cite{bayer1997amyloid}\\} \\
\hline
CD24 & \makecell{- Over-expression stimulates glioma cell migration, \\ invasion, colony formation, and tumor growth in mice} & \makecell{\cite{barash2019heparanase}} \\
\hline
LAPTM5 & \makecell{- May be used by glioma to negatively \\ modulate antitumoral response of immune system} & \makecell{\cite{m2020aberrant}} \\
\hline
C1QB & \makecell{- Upregulated in diverse grade of gliomas \\ and plays important role in pathogenesis} & \makecell{\cite{mangogna2019prognostic}} \\
\hline
IL1B & \makecell{- Highly expressed in GBM and \\ promotes glioma spheroid formation} & \makecell{\cite{wang2019ccaat}} \\
\hline
\end{longtable}
\end{center}
\normalsize	
We see in Table~\ref{tab:imp} that PLP1, BCAN, and PTPRZ1 are all associated with the development and malignant growth of glioma. This is a striking observation since, in terms of gene importance score, three of the seven top differentially expressed genes play notable roles in glioma progression and malignancy. Along with this consideration, we observe that a large proportion of the top forty important genes are associated with brain cancer and neurological or neurodegenerative disease. This shows that RZiMM-scRNA is quite effective in detecting biologically important genes in real scRNA-seq data.

It is also noteworthy that many of the top over-expressed genes in each cluster established by our new method serve similar biological functions. Upon evaluation of cluster-specific expression difference (\ref{diff}), we determine the top over- and under-expressed genes in specific cell clusters. For example, we find that the top upregulated genes in Cluster 1 are RGS1, CD74, CCL3, CCL4, and HLA-DRA. We see in Table S2 that these genes are involved in immune function and regulation, revealing that particular cells dealing with immune system mediation in glioma are grouped together by RZiMM-scRNA. Likewise, we observe that PLP1, CLDN11, MOG, and MBP are all highly over-expressed in Cluster 2 and note that these genes play roles in oligodendrocyte development and myelination. The fact that the top upregulated genes in each cluster serve similar biological functions suggests that RZiMM-scRNA is effective in clustering individual cells into groups in accordance with their biological functions.

In addition to new biomarkers discovered by our method, our findings are also supported by the previous analysis done by \citeauthor{venteicher2017decoupling}. They reported the presence of an inflammatory program active in GAMs consisting of cytokines and chemokines, in which genes such as CCL3, CCL4, and IL1B are active. This is consistent with our result that these genes, along with other genes involved in immune regulation, are over-expressed in Cluster 1. Likewise, the previous analysis discovered a group of nonmalignant oligodendrocytes in their clustering that preferentially expressed markers such as MBP and CLDN11. These genes are also upregulated in our Cluster 2 along with other oligodendrocytic markers. Moreover, the efficacy of our method is proven by our finding that the most under-expressed genes across these two cell groups include BCAN, PTPRZ1, and GPM6A. Since these genes are involved in neuronal migration and the growth of brain tumor cells, the downregulation of these genes is consistent with these cells being nonmalignant, which is confirmed by the properties we discuss.

\subsection{Analysis of High-Grade Astrocytoma Data}

In this study of aggressive brain tumor types, conducted by \citeauthor{yuan2018single}, samples from eight patients were taken of the cellular milieu of high-grade gliomas, namely glioblastoma and anaplastic astrocytoma. Taken together, $N = 23,793$ total cells were profiled across $J = 60,725$ total genes, among which the majority have very sparse data. Posterior to the similar filtering procedure conducted in the first study, we retain 3550 genes in our data analysis.


Based on the empirical distribution of this dataset, a zero-inflated Poisson distribution seems to provide better fit than the Gaussian distribution. We thus implement our algorithm assuming the zero-inflated Poisson distribution, in which the number of cell subgroups is fixed to six, as suggested in the original study by \citeauthor{yuan2018single} for fair comparisons.  



Out of the top $40$ genes identified by our method, $16$ of these are important genes found in the previous analysis---KLK4, KIF18B, NCAPH, NCAPG, SUSD3, TROAP, ABCB1, MYBPC1, KIFC1, GTSE1, PRC1, MAD2L1, TPX2, and EFHD1. 
This finding helps support the meaningfulness of our obtained result. 
In addition, our method demonstrates its potential to make new discoveries as we continue to investigate the biological significance of the top genes which are not  discussed in the previous work. Excitingly, we find that many of these genes are associated with cancers as well as neurological and neurodegenerative diseases. We include a brief description of the biological significance of some of these genes in Table~\ref{tab:imp2}. 

\footnotesize
\begin{center}
\setlength\arrayrulewidth{0.01pt}
\begin{longtable}{|c|c|c|}
\caption{Top Genes Not Previously Discussed}\label{tab:imp2}\\
\hline
\textbf{Gene} & \textbf{Significance} & \textbf{Ref.} \\
\hline
\endfirsthead
\multicolumn{3}{c}%
{\tablename\ \thetable\ -- \textit{Continued from previous page}} \\
\hline
\textbf{Gene} & \textbf{Significance} & \textbf{Ref.} \\
\hline
\endhead
\hline \multicolumn{3}{r}{\textit{Continued on next page}} \\
\endfoot
\hline
\endlastfoot
CAMK2B & \makecell{- Downregulated in GBM and glioma, \\ implicated in neurodevelopment, brain \\ function, learning and memory processes} &
\makecell{\cite{xiong2019silico} \\ \cite{johansson2005expression} \\ \cite{nicole2020camkiibeta}} \\
\hline
LZTS2 & \makecell{- Tumor suppressor gene influencing \\ proliferation of cancer cell lines such \\ as breast, colon, prostate, and glioma} & \makecell{\cite{thyssen2006lzts2} \\ \cite{kim2011effect}} \\
\hline
CBR1 & \makecell{- Decreased expression associated \\ with tumor metastasis and poor prognosis \\
- Upregulated in AD brains \\
- Partially responsible for down \\ syndrome cognitive impairments} & 
\makecell{ \cite{murakami2011suppression} \\ \cite{balcz2001increased} \\ \cite{arima2020impairment}} \\
\hline
PAQR8 & \makecell{- Downregulated in GBM and \\ malignant endometrial cancer tissue} & 
\makecell{\cite{moral2020role} \\ \cite{sinreih2018membrane}} \\
\hline
KANK2 & \makecell{- Potential tumor suppressor gene \\ involved in regulation of apoptosis, \\ cell cycle, and transcription} & \makecell{\cite{zhang2018genome} \\ \cite{wang2014chromosome}} \\
\hline
ZNF713 & \makecell{- Misregulation in brain may be involved \\ in pathogenicity of autistic disorder} & \makecell{\cite{metsu2014cgg}} \\
\hline
USP54 & \makecell{- Overexpressed in colorectal cancer stem \\ cells and promotes intestinal tumorigenesis} & \makecell{\cite{fraile2016deubiquitinase}} \\
\hline
MSANTD3 & \makecell{- May be associated with familial \\ squamous cell lung carcinoma} & \makecell{ \cite{byun2018genome}} \\
\hline
DYNLL1P1 & \makecell{- May be associated with schizophrenia} & \makecell{\cite{peyrot2021identifying}} \\
\hline
\end{longtable}
\end{center}
\normalsize	

We observe in Table~\ref{tab:imp2} that many of our identified new biomarkers are connected to tumorigenicity, neurodevelopment, and neurological diseases. For example, CAMK2B is found to be downregulated in GBM and is implicated in neuro-development, brain function, learning and memory processes. Likwise, we find that PAQR8 is downregulated in GBM and malignant endometrial cancer tissues. The application of our method to this high-grade astrocytoma data further demonstrates the effectiveness of RZiMM-scRNA in detecting biologically meaningful genes in real scRNA-seq data.


\section{Discussion}

Single-cell RNA-seq is a valuable technique for gaining important biological insight into the formation and progression of heterogeneous diseases. However, using inappropriate methods to analyze such data can result in incomplete or misleading biological conclusions. Compared to bulk sample RNA-seq, the data generated from scRNA-seq possess unique characteristics, including the presence of dropouts and batch/sample effects, which brings additional challenges into data analysis. Without proper treatment for these issues, biologically meaningful discoveries, such as underlying cell heterogeneity patterns, cannot be made by unsupervised clustering algorithms. In this paper, we present a novel analytical method --- RZiMM-scRNA, a unified statistical framework to simultaneously discover cell subtypes and identify differential gene expression, while accounting for both dropouts and sample effects. We illustrate its superior performance in terms of both clustering and biomarker identification over several widely used methods including Seurat, SC3, K-Means, H-Clust, and fastMNN, via comprehensive simulation studies in some commmon settings and an application to astrocytoma scRNA-seq data.

In the simulation studies, RZiMM-scRNA demonstrates its effectiveness in discovering hidden cell subgroups and detecting important biomarkers, despite the presence of dropout events and sample effects. As demonstrated by the count data simulation study, our new framework can be straightforwardly implemented under non-Gaussian distributions, for example, the zero-inflated Poisson or negative binomial models that are widely used for the count data. Interestingly, we also observe that the performance of adopting zero-inflated Gaussian distribution in the proposed framework is quite robust to different underlying distribution forms that are widely considered for scRNA-Seq data. 

Moreover, we find during the first real astrocytoma scRNA-seq data application that RZiMM-scRNA identifies two biologically meaningful cell types, previously annotated as microglia and oligodendrocytes by \citeauthor{venteicher2017decoupling}, while the other methods fail to identify the small oligodendrocyte type. Utilizing the model framework of RZiMM-scRNA, we are able to identify top genes that are significantly over-expressed in the oligodendrocyte type, some of which are also reported by \citeauthor{venteicher2017decoupling}. The informative paradigm of RZiMM-scRNA also allows us to identify top genes which are differentially expressed across all the hidden cell subgroups, most of which are confirmed to be GBM-related by the related literature. Likewise, the second real data application to high-grade astrocytoma tumors further demonstrates the promise of our model framework in terms of biomarker detection. 
 
In our future investigation, we will develop bootstrapping methods to assess the statistical significance of identified genes that differentiate between discovered cell subtypes.

\begin{funding}
The work was partially supported by the National Institute of Health Grants NCI 5P30 CA013696, NCI P01 CA098101, NIAID 1R01 AI143886, and NCI 1R01 CA219896. 
\end{funding}

\begin{supplement}
\stitle{Supplementary Information}
\sdescription{We provide additional plots and tables for the results section in Supplementary Information. }
\end{supplement}
\begin{supplement}
\stitle{Code and Software}
\sdescription{We have developed an R package ``\href{https://github.com/SkadiEye/RZiMM}{RZiMM}'', to implement our proposed method, and scripts to perform simulational and real applications. }
\end{supplement}


\bibliographystyle{imsart-nameyear} 
\bibliography{references}       

\end{document}


\maketitle

\begin{figure}
    \centering
    \includegraphics[width = \textwidth]{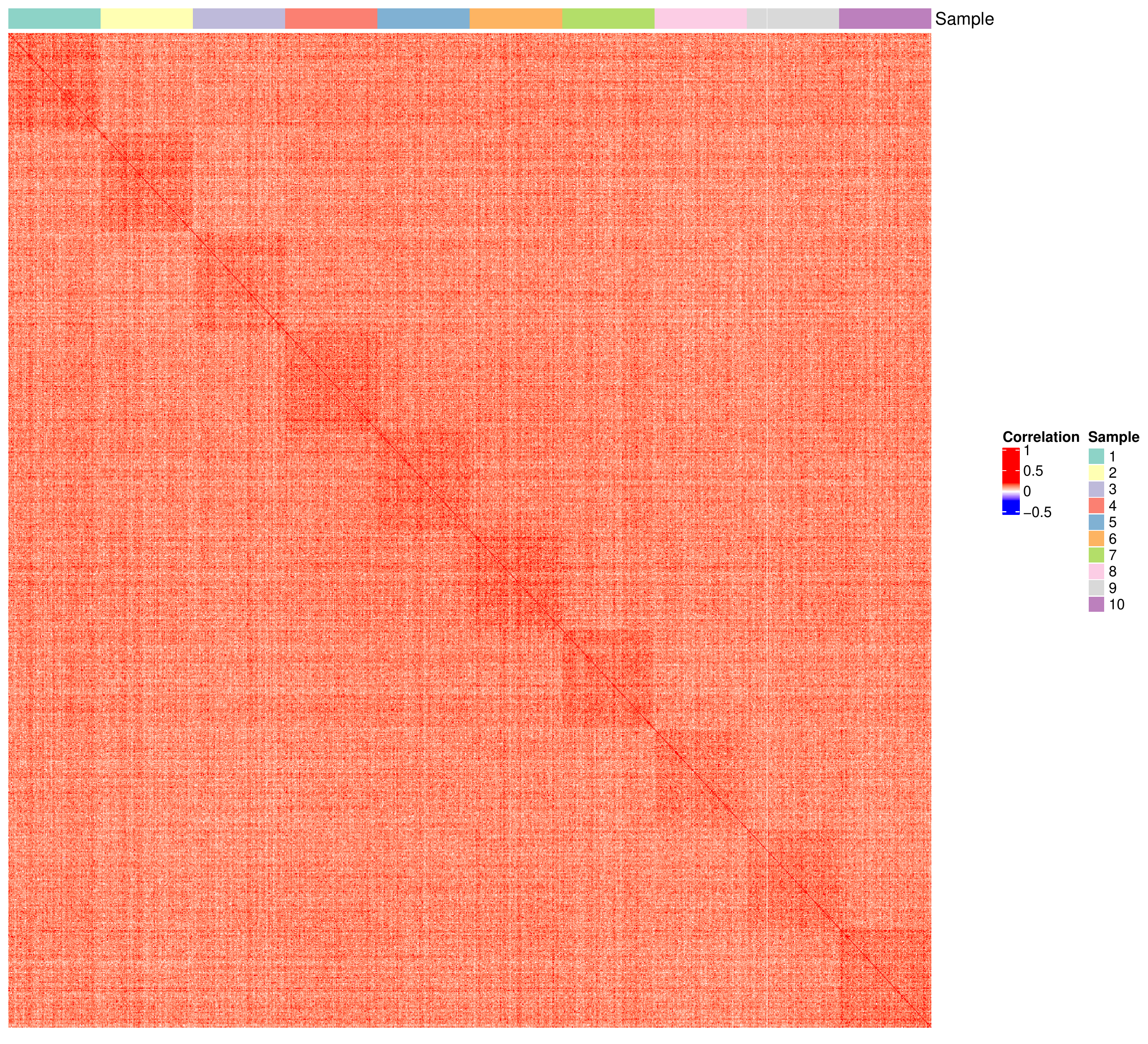}
    \caption{Correlation structure for the simulation study at $\delta = 0.5$}
    \label{fig:sim}
\end{figure}

\begin{figure}
    \centering
    \includegraphics[width = \textwidth]{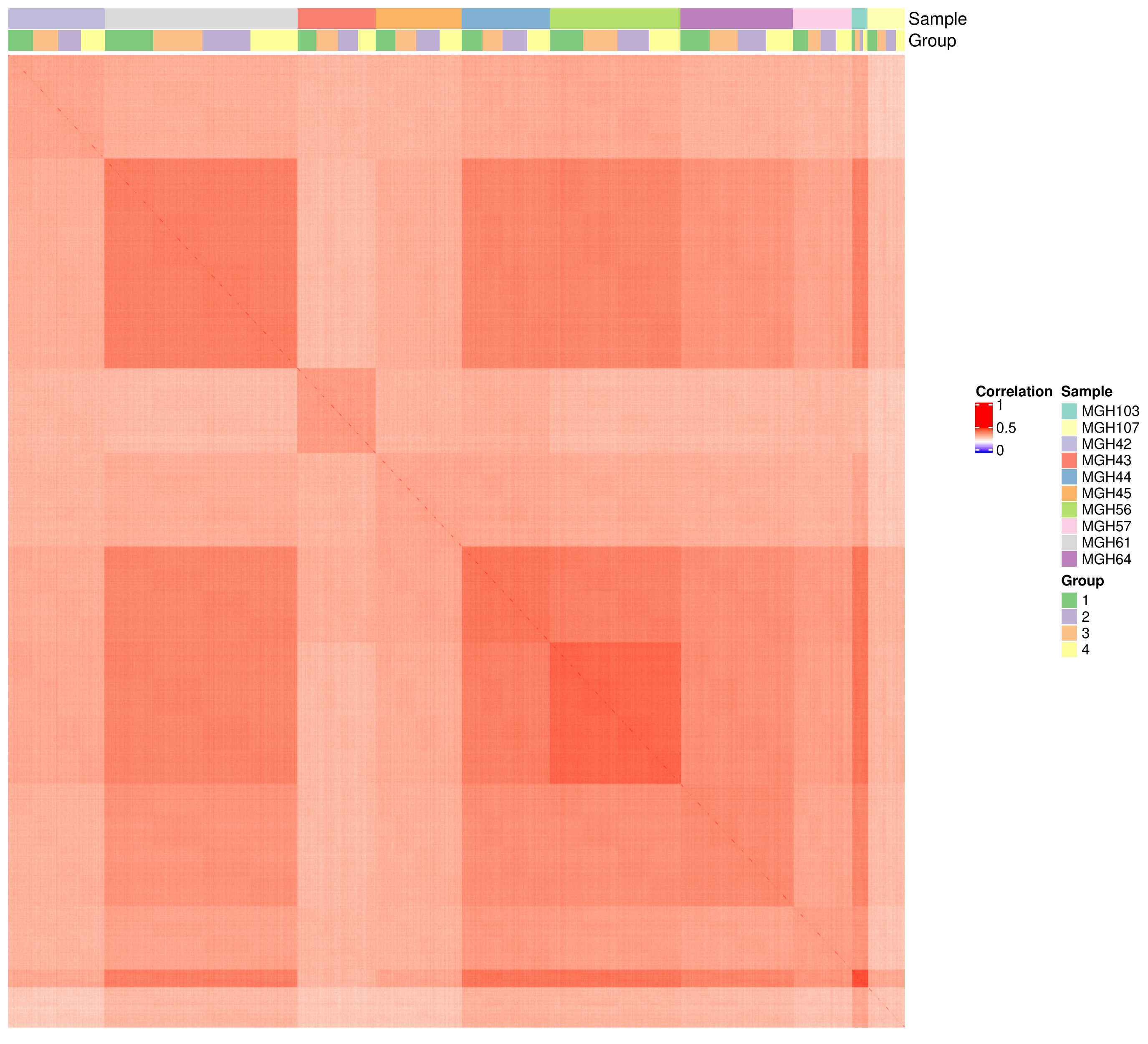}
    \caption{Correlation structure for the semi-simulation study at $v = 4$}
    \label{fig:semi_sim}
\end{figure}

\begin{figure}
    \centering
    \includegraphics[width = \textwidth]{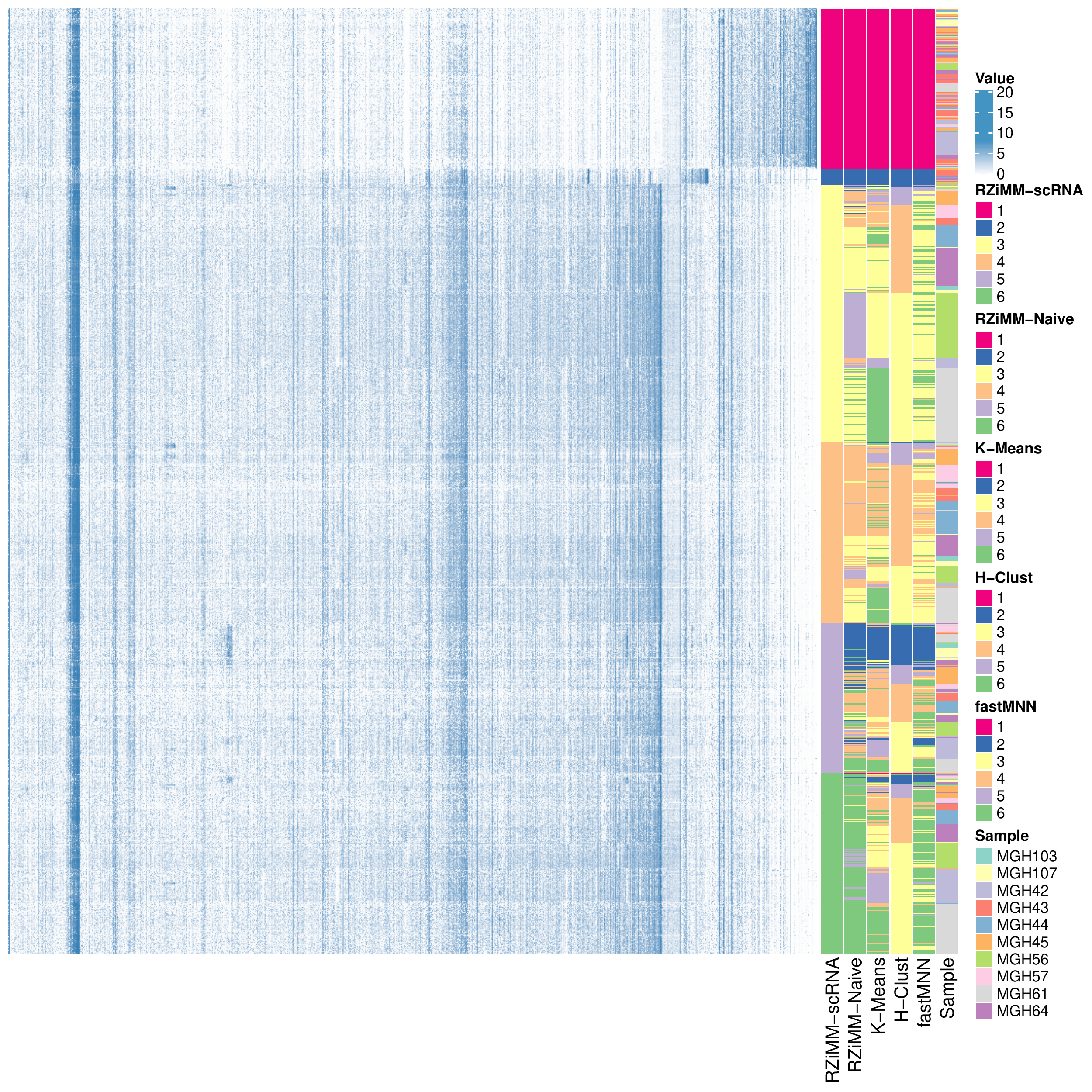}
    \caption{Heatmap for glioma scRNA-seq data. Cells are ordered by RZiMM-scRNA and H-Clust; genes are ordered by H-Clust. }
    \label{fig:heatmap}
\end{figure}

\begin{figure}
    \centering
    \includegraphics[width = \textwidth]{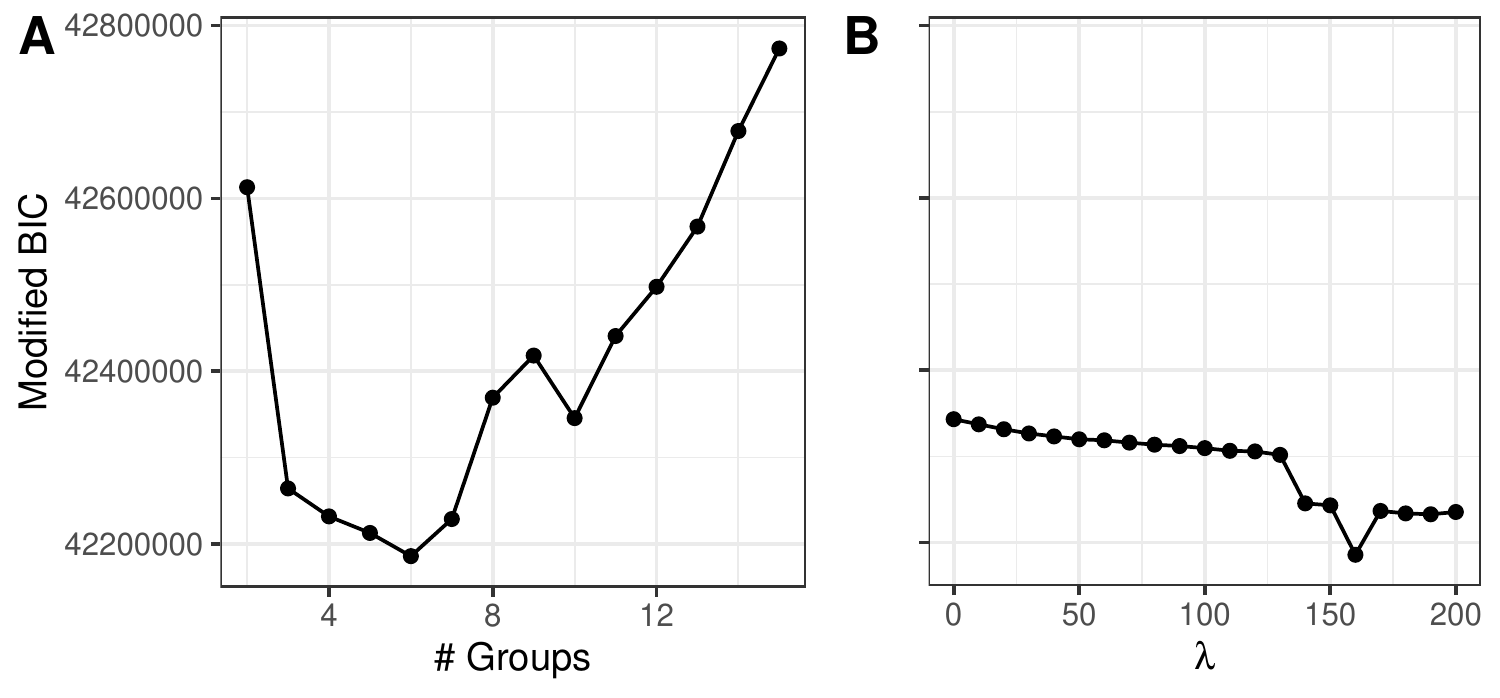}
    \caption{Modified BIC for RZiMM-scRNA }
    \label{fig:bic}
\end{figure}

\begin{figure}
    \centering
    \includegraphics[width = \textwidth]{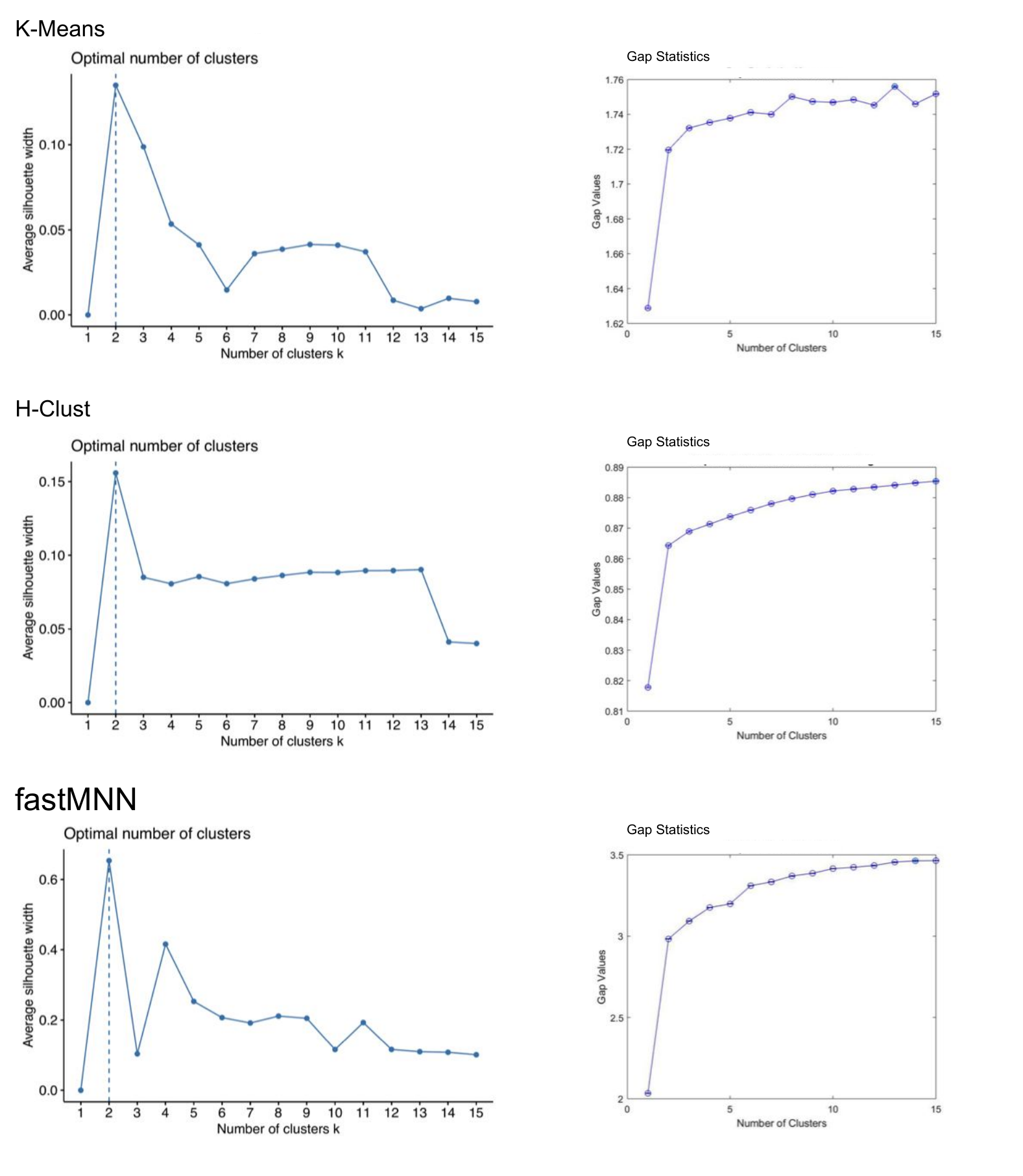}
    \caption{Selecting optimal numbers of clusters for K-Means, H-Clust and fastMNN. From top to bottom: K-Means, H-Clust, fastMNN; from left to right: the average silhoutte value method and gap statistics. }
    \label{fig:sil}
\end{figure}

\FloatBarrier
\begin{table}[ht]
\centering
\setstretch{1.2}
\begin{tabular}{l|r|rrrrrr|l}
  \hline
Gene & Imp. & $m_{j1}$ & $m_{j2}$ & $m_{j3}$ & $m_{j4}$ & $m_{j5}$ & $m_{j6}$ & \multicolumn{1}{c}{Note} \\ 
  \hline
SPP1 & 1.81 & 9.72 & 8.95 & 2.73 & 2.73 & 2.58 & 2.75 &   OE in C1, C2 \\ 
  PLP1 & 1.68 & 2.67 & 12.11 & 4.42 & 4.40 & 5.31 & 4.93 &   UE in C1; OE in C2 \\ 
  BCAN & 1.61 & 2.22 & 4.23 & 9.07 & 8.84 & 8.81 & 8.88 &   UE in C1, C2 \\ 
  RGS1 & 1.43 & 10.17 & 2.56 & 2.02 & 2.02 & 2.02 & 2.56 &   OE in C1 \\ 
  CD74 & 1.43 & 10.28 & 2.74 & 2.14 & 2.29 & 2.29 & 2.82 &   OE in C1 \\ 
  GPM6A & 1.41 & 1.79 & 3.48 & 7.60 & 7.08 & 7.58 & 7.70 &   UE in C1, C2 \\ 
  PTPRZ1 & 1.33 & 1.61 & 2.77 & 7.12 & 6.56 & 6.60 & 6.91 &   UE in C1, C2 \\ 
  CCL3 & 1.29 & 9.27 & 1.53 & 1.53 & 1.53 & 1.53 & 1.53 &   OE in C1 \\ 
  SEPP1 & 1.28 & 4.86 & 8.93 & 2.67 & 2.67 & 4.44 & 3.86 &   OE in C2 \\ 
  GSN & 1.27 & 6.86 & 8.08 & 2.89 & 2.89 & 3.62 & 3.36 &   OE in C1, C2 \\ 
  FXYD6 & 1.23 & 2.67 & 4.49 & 7.95 & 7.93 & 7.62 & 7.83 &   OE in C1, C2 \\ 
  MOG & 1.21 & 0.88 & 7.65 & 0.92 & 0.73 & 0.84 & 1.40 &   OE in C2 \\ 
  MBP & 1.21 & 3.26 & 10.07 & 3.39 & 3.38 & 3.95 & 3.95 &   OE in C2 \\ 
  CCL4 & 1.21 & 8.94 & 1.71 & 1.71 & 1.71 & 1.71 & 1.71 &   OE in C1 \\ 
  A2M & 1.21 & 7.60 & 1.01 & 0.61 & 0.63 & 1.01 & 1.02 &   OE in C1 \\ 
  APOE & 1.21 & 7.81 & 3.27 & 3.14 & 3.14 & 6.76 & 5.23 &   OE in C1 \\ 
  CRYAB & 1.20 & 3.18 & 9.20 & 4.02 & 3.94 & 5.70 & 4.62 &   OE in C2 \\ 
  CLDN11 & 1.19 & 0.58 & 7.47 & 0.58 & 0.53 & 0.56 & 0.88 &   OE in C2 \\ 
  FEZ1 & 1.19 & 0.74 & 7.52 & 4.23 & 4.15 & 4.64 & 4.52 &   UE in C1; OE in C2 \\ 
  HLA-E & 1.16 & 6.79 & 2.42 & 0.93 & 0.76 & 1.69 & 1.74 &   OE in C1 \\ 
  TF & 1.13 & 2.09 & 8.43 & 2.23 & 2.07 & 2.64 & 2.65 &   OE in C2 \\ 
  HLA-DRA & 1.13 & 9.65 & 2.88 & 2.88 & 2.88 & 2.88 & 2.88 &   OE in C1 \\ 
  C1orf61 & 1.12 & 2.37 & 3.96 & 6.73 & 6.80 & 7.36 & 6.72 &   UE in C1, C2 \\ 
  APLP1 & 1.10 & 1.08 & 7.14 & 2.62 & 2.38 & 3.11 & 3.16 &   UE in C1; OE in C2 \\ 
  CD24 & 1.09 & 0.72 & 1.26 & 4.90 & 4.68 & 3.58 & 4.84 &   UE in C1, C2 \\ 
  LAPTM5 & 1.09 & 8.11 & 2.34 & 1.91 & 1.91 & 1.91 & 2.34 &   OE in C1 \\ 
  C1QB & 1.08 & 8.90 & 2.40 & 2.40 & 2.40 & 2.40 & 2.40 &   OE in C1 \\ 
  IL1B & 1.08 & 5.50 & -0.99 & -0.99 & -0.99 & -0.99 & -0.99 &   OE in C1 \\ 
  GPM6B & 1.08 & 2.15 & 8.37 & 6.77 & 6.65 & 7.00 & 6.66 &   UE in C1, OE in C2 \\ 
  MEG3 & 1.06 & -0.22 & 0.29 & 3.57 & 3.09 & 3.38 & 4.10 &   UE in C1, C2 \\ 
  SRGN & 1.05 & 8.33 & 2.07 & 2.07 & 2.07 & 2.07 & 2.18 &   OE in C1 \\ 
  CCL4L1 & 1.05 & 6.50 & 0.22 & 0.22 & 0.22 & 0.22 & 0.22 &   OE in C1 \\ 
  CCL4L2 & 1.04 & 6.50 & 0.22 & 0.22 & 0.22 & 0.22 & 0.22 &   OE in C1 \\ 
  TUBA1A & 1.04 & 4.56 & 10.68 & 9.41 & 9.32 & 9.17 & 9.33 &   UE in C1; OE in C2 \\ 
  DBNDD2 & 1.02 & 4.31 & 9.31 & 3.61 & 3.68 & 4.27 & 3.96 &   OE in C2 \\ 
  C3 & 1.02 & 7.33 & 1.19 & 1.19 & 1.19 & 1.19 & 1.19 &   OE in C1 \\ 
  FKBP5 & 1.02 & 5.15 & 2.54 & 0.40 & 0.34 & 0.42 & 0.61 &   OE in C1, C2 \\ 
  CSF1R & 1.02 & 7.45 & 1.32 & 1.32 & 1.32 & 1.32 & 1.32 &   OE in C1 \\ 
  PMP2 & 1.02 & 3.75 & 5.41 & 7.99 & 7.43 & 8.18 & 8.02 &   UE in C1, C2 \\ 
  HLA-B & 1.01 & 7.27 & 4.59 & 2.48 & 2.53 & 3.30 & 3.21 &   OE in C1, C2 \\ 
   \hline
\end{tabular}
\caption{Top 40 Significant Genes Identified by RZiMM-scRNA. Imp.: Importance; OE: over-expressed; UE: under-expressed; C1, C2: Cluster 1, Cluster 2. }
\end{table}

\FloatBarrier


\setstretch{1.4}
\begin{center}
\begin{longtable}{|c|c|c|}
\caption{Function and Significance of Top 40 Important Genes}\\
\hline
\textbf{Gene} & \textbf{Function \& Significance} & \textbf{Ref.} \\
\hline
\endfirsthead
\multicolumn{3}{c}%
{\tablename\ \thetable\ -- \textit{Continued from previous page}} \\
\hline
\textbf{Gene} & \textbf{Function \& Significance} & \textbf{Ref.} \\
\hline
\endhead
\hline \multicolumn{3}{r}{\textit{Continued on next page}} \\
\endfoot
\hline
\endlastfoot
SPP1 & \makecell{- Functions in immune regulation, tumorigenesis, and cell signaling \\ - Regulates tumor progression by increasing \\ migration, invasion and cancer stem cell self-renewal \\ - Upregulated by GAMs and associated \\ with poor patient prognosis in LGG and GBM\\ - Active in PI3K-Akt signaling pathway, which regulates \\ cellular functions such as proliferation, growth, and survival\\ - Active in human papillomavirus infection pathway} &
\makecell{\citep{chen2019identification} \\ \citep{kijewska2017embryonic}\\ \\
\citep{chen2019identification,szulzewsky2015glioma} \\ \\
\citep{kanehisa2000kegg}\\\\
\citep{kanehisa2000kegg}\\} \\
\hline
PLP1 & \makecell{- Involved in compaction, stabilization, and maintenance of \\ myelin sheaths, oligodendrocyte development, and axonal survival \\ - Key for glioma development and associated with patient prognosis \\ - Downregulated across variety of cancers, such as GBM, prostate  \\ adenocarcinoma, colon adenocarcinoma, and thyroid carcinoma. \\ - Mutations cause neurological diseases, such as X-linked \\ Pelizaeus-Merzbacher disease and Spastic Paraplegia type 2 \\ - Mutations could induce MS in a small proportion of patients} & \makecell{\citep{o2016reference} \\\\ \citep{chen2019screening}\\
\citep{li2017transcriptional}\\\\
\citep{margraf2018novel}\\\\
\citep{cloake2018plp1}\\} \\
\hline
BCAN & \makecell{- May promote the growth and motility of brain tumor cells \\ - Overexpressed in grade II/III glioma and secondary GBM} & \makecell{\citep{o2016reference} \\ \citep{dutoit2018antigenic,ma2012associations}} \\
\hline
RGS1 & \makecell{- Negatively regulates chemokine receptor signaling in lymphocytes \\ - Gene polymorphisms significantly associated with MS \\ - High expression associated with poor prognosis in multiple myeloma \\ - Associated with etiology and prognosis of \\ melanoma and late-stage non-small cell lung cancer } & 
\makecell{\citep{caballero2016autoimmunity} \\ \citep{caballero2016autoimmunity,sawcer2011genetic}\\
\citep{roh2017rgs1}\\
\citep{rangel2008novel,dai2011genetic}\\\\} \\
\hline
CD74 & \makecell{- Serves as receptor for macrophage migration inhibitory factor \\ which functions in innate and adaptive immunity, and cell proliferation \\ - Upregulated in HGG and positively associated with patient survival\\ - Overexpressed in various neoplasms, mainly in hematologic tumors} & \makecell{\citep{leng2003mif} \\\\ \citep{zeiner2015mif}\\ \citep{zeiner2015mif}} \\
\hline
GPM6A & \makecell{- Implicated in differentiation and migration of neurons \\ from human stem cells, and filopodium outgrowth and motility\\ - Downregulated in the hippocampus of chronically stressed animals \\ - Transcriptional downregulation in hippocampus of depressed suicides \\ - Identified as a top gene candidate as a causal gene for SZ \\ - Overexpressed in mantel cell and chronic lymphocytic leukemia} & \makecell{\citep{fuchsova2015altered,michibata2009human} \\ \citep{fuchsova2009cysteine} \\ \citep{fuchsova2015altered}\\
\citep{fuchsova2015altered}\\
\citep{ma2018integrated}\\
\citep{charfi2014identification} \\} \\
\hline
PTPRZ1 & \makecell{- Required for normal differentiation of precursor cells \\ into mature, fully myelinating oligodendrocytes, and may \\ play a role in protecting oligodendrocytes against apoptosis \\ - Involved in tumor cell proliferation and tumor growth\\ - Upregulated in LGG and GBM, and associated with \\ malignant growth of GBM and poor patient prognosis\\ - Implicated in function of prefrontal cortex, a key \\ area in depression, anxiety-related disorders, and SZ \\ - Gene polymorphisms significantly associated with SZ susceptibility} & \makecell{\citep{uniprot2019uniprot} \\\\\\ \citep{xia2019expression}\\
\citep{xia2019expression,shi2017tumour}\\\\
\citep{cressant2017loss}\\\\
\citep{cressant2017loss,buxbaum2008molecular}\\} \\
\hline
CCL3 & \makecell{- Chemokine involved in recruitment and activation of granulocytes \\ - Downregulated in HGG \\ - Patients with higher expression levels have better survival \\ - Highly expressed in AD mouse model} & \makecell{\citep{o2016reference,wolpe1988macrophages} \\ \citep{vidyarthi2019predominance}\\
\citep{vidyarthi2019predominance} \\
\citep{jorda2019changes}} \\
\hline
SEPP1 & \makecell{- Transports selenium and acts as extracellular antioxidant \\ - Inactivation causes severe neurological dysfunction \\ and brainstem neurodegeneration in mice \\ - Elevated expression in AD brain \\ - Decreased expression in substantia nigra of PD subjects \\ - Downregulated in variety of cancers, such as renal cell carcinoma, \\ colorectal adenoma, prostate cancer, and gastric adenocarcinoma
} & \makecell{\citep{o2016reference} \\ \citep{byrns2014mice}\\\\
\citep{rueli2015increased}\\
\citep{bellinger2012changes}\\
\citep{rizk2019hmgb1,guariniello2015evaluation}\\\\} \\
\hline
GSN & \makecell{- Regulates cell morphology, differentiation, movement, and apoptosis \\ - Restrains invasion and migration of colon cancer cells \\ - Significantly downregulated in many cancers, such as colon and GBM\\ - Involved in AD pathophysiology and \\ differentially expressed in urine of AD patients \\ - Decreased secreted GSN expression in blood and \\ increased expression in brain detected in MS mouse model \\ - Plays role in the pathway for viral carcinogenesis} & \makecell{\citep{hu2014gelsolin,silacci2004gelsolin} \\ \citep{chen2019lower}\\
\citep{chen2019lower,miyauchi2018identification}\\
\citep{yao2018urine}\\\\
\citep{li2015gelsolin}\\\\
\citep{kanehisa2000kegg}\\
} \\
\hline
FXYD6 & \makecell{- Regulates the activity of Na$^+$/K$^+$-ATPase, \\ which is involved in tumor proliferation and migration \\ - Upregulated in hepatocellular carcinoma (HCC) \\ and enhances  HCC cell migration and proliferation \\ - Upregulated in human osteosarcoma cells \\ - Downregulated in brain of AD mouse model \\ - Gene mutations are likely to increase genetic susceptibility to SZ} & \makecell{\citep{gao2014fxyd6,prassas2008novel} \\\\ \citep{gao2014fxyd6}\\ \\
\citep{olstad2003molecular}\\
\citep{george2010serial}\\
\citep{choudhury2007genetic}} \\
\hline
MOG & \makecell{- Glycoprotein potentially involved in maintaining \\ structural integrity of the myelin sheath in CNS \\ - Anti-MOG antibodies associated with demyelinating diseases, \\ such as acute disseminated encephalomyelitis, neuromyelitis optica \\ spectrum disorder, and chronic relapsing inflammatory optic neuropathy} & \makecell{\citep{roth1995human} \\\\ \citep{sinmaz2016mapping}\\\\\\} \\
\hline
MBP & \makecell{- Major constituent of myelin sheath of oligodendrocytes \\ and active in the myelination of nerves in the NS \\ - Increased expression levels detected in AD cortex \\ and could be involved in amyloid plaque formation \\ - Elevated expression levels in CSF of patients with progressive MS \\ - Increased expression levels in CSF after severe pediatric TBI} & \makecell{\citep{o2016reference,uniprot2019uniprot}\\\\ \citep{zhan2015myelin}\\\\
\citep{sellebjerg2017defining}\\
\citep{su2012increased}\\} \\
\hline
CCL4 & \makecell{- Acts as chemoattractant for immune cells \\ including natural killer cells and monocytes \\ - Upregulated by GAMs \\ - Functions in NF-$\kappa$B signaling pathway, \\ often misregulated in tumorigenic cells} & \makecell{\citep{bystry2001b} \\\\ \citep{zeiner2019distribution}\\ \citep{kanehisa2000kegg}\\\\} \\
\hline
A2M & \makecell{- Inhibits a broad spectrum of proteases and inflammatory cytokines \\ - Inhibits proliferation, migration, and invasion of astrocytoma cells \\ - May contribute to pathogenesis of AD} & \makecell{\citep{o2016reference} \\ \citep{lindner2010alpha2}\\ \citep{blacker1998alpha}} \\
\hline
APOE & \makecell{- Acts in lipid transport in CNS, regulating neuron survival and sprouting \\ - Upregulated in grade II-IV astrocytoma \\ - Involved in $A\beta$ aggregation in pathogenesis of AD \\ - Polymorphism associated with neurological diseases and psychiatric \\ disorders, such as AD, SZ, MS, PD, brain microbleeds, and dementia \\ - Anti-angiogenic and metastasis-suppressive factor} & 
\makecell{\citep{uniprot2019uniprot} \\
\citep{rousseau2006expression}\\
\citep{kanehisa2000kegg}\\
\citep{forero2018apoe}\\\\
\citep{pencheva2012convergent}\\} \\
\hline
CRYAB & \makecell{- Involved in chaperone-like, autokinase, and anti-apoptotic activity \\ - Upregulated in IDH1$^\text{R132H}$ mutant anaplastic astrocytoma and GBM \\ - Associated with lymph node metastasis, tumor differentiation, \\ invasion depth, and unfavorable outcomes in human gastric cancer \\ - Dysfunction closely related to AD, PD, and Creutzfeldt-Jakob disease \\ - Downregulated in HD mouse model \\ and contributes to its neuropathology} & \makecell{\citep{o2016reference,lee2012expression} \\
\citep{avliyakulov2014c,kore2014inflammatory} \\
\citep{tao2019expression}\\\\
\citep{liu2017structural}\\
\citep{hong2017mutant}\\\\} \\
\hline
CLDN11 & \makecell{- Regulates proliferation and migration of oligodendrocytes \\ - Absence in CNS myelin perturbs behavior in \\ mice and could lead to neuropsychiatric disease \\ - Reduced expression in SZ brain \\ - Overexpression inhibits proliferation,\\ migration, and invasion of gastric carcinoma cells \\ - Downregulation promotes development of gastric cancer
} & \makecell{\citep{o2016reference} \\
\citep{maheras2018absence} \\\\
\citep{dracheva2006myelin}\\
\citep{yang2017microrna}\\\\
\citep{yang2017microrna}\\} \\
\hline
FEZ1 & \makecell{- Involved in axonal outgrowth and elongation \\ - Abnormal, unphosphorylated co-aggregates \\ present in brains of transgenic AD mice models \\ - Involved in astrocytic protection of dopamine neurons and \\ regulation of  neuronal microenvironment during PD progression
 \\ - Tumor suppressor gene inactivated in many cancers, \\ such as prostate, esophageal, gastric, bladder, and breast \\ - Negatively associated with tumor grading and mortality rate\\} & \makecell{\citep{o2016reference} \\
\citep{butkevich2016phosphorylation}\\\\
\citep{sun2014fasciculation}\\ \\ 
\citep{nonaka2005reduced}\\ \\
\citep{nonaka2005reduced}\\
} \\
\hline
HLA-E & \makecell{- Functions in cell recognition by natural \\ killer cells and regulation of cell-mediated lysis \\ - Upregulated in grade II-IV astrocytoma\\ - Correlated with more aggressive tumor \\  grade and histological type in diffuse glioma\\  - Upregulated in MS active lesions \\ -  Involved with neuroactive ligand-receptor interaction, \\ associated with SZ, ADHD, alcohol dependence, and neurosis \\ - Involved in Epstein-Barr virus (EBV) infection pathway, which is \\ linked to nasopharyngeal carcinoma, Hodgkin's, and Burkitt's lymphoma \\ - Plays role in human T-cell leukemia virus 1 infection\\ pathway, which is associated with adult T-cell leukemia/lymphoma} & \makecell{\citep{uniprot2019uniprot} \\\\ \citep{mittelbronn2007elevated}\\
\citep{wu2020hla}\\\\
\citep{durrenberger2012increased}\\
\citep{kanehisa2000kegg}\\ \\
\citep{kanehisa2000kegg}\\\\
\citep{kanehisa2000kegg}\\\\} \\
\hline
TF & \makecell{- Plays central role in ferric-ion delivery to proliferating body cells \\ - Upregulated on surface of cancer cells due to elevated iron demand \\ - Free transferrin in blood may competitively bind to \\ malignant cell  receptors, preventing uptake and limiting efficacy \\ - May play role in increased iron deposition in MS brain} & \makecell{\citep{o2016reference} \\ \citep{oliveira2017targeted}\\
\citep{alexander2013bioengineering,xu2001systemic}\\\\
\citep{khalil2014cerebrospinal}\\} \\
\hline
HLA-DRA & \makecell{- Presents extracellular peptide antigens to immune system to \\ elicit or suppress T-cell responses that lead to antibody production \\ - Low expression levels associated with poor \\ prognosis in pediatric adrenocortical tumors \\ - Significantly downregulated in osteosarcoma \\ - Involved in EBV infection pathway} & \makecell{\citep{uniprot2019uniprot} \\ \\ \citep{leite2014low}\\\\
\citep{pan2018identification}\\
\citep{kanehisa2000kegg}\\} \\
\hline
C1orf61 & \makecell{- Involved in development and remodeling of neurons \\ - Acts as brain-specific transcriptional activator of \\ human c-fos proto-oncogene, a regulator of neuronal death \\ pathways associated with cell proliferation and differentiation \\ - May participate in occurrence and development of GBM \\ - Associated with progression of liver disease \\ and promotes HCC growth and metastasis \\} & \makecell{\citep{jeffrey2000croc} \\ \citep{jeffrey2000croc}\\\\\\ \citep{jia2019integrative}\\ \citep{hu2013c1orf61}\\\\} \\
\hline
APLP1 & \makecell{- Involved in synaptic maturation and regulation of neurite outgrowth \\ - Upregulated during retinoic acid-induced \\ differentiation of human neuroblastoma cells \\ - May contribute to pathogenesis of AD-associated neurodegeneration \\ - Overexpression associated with poor overall \\ survival in early stage clear cell renal cell carcinoma} & \makecell{\citep{o2016reference,uniprot2019uniprot} \\ \citep{beckman1997increased}\\\\
\citep{bayer1997amyloid}\\
\citep{batai2018whole}\\\\} \\
\hline
CD24 & \makecell{- Mediates neuronal proliferation, \\ differentiation, and immune suppression in CNS \\ - Upregulated in anaplastic astrocytoma, GBM, and medulloblastoma \\ - Overexpression stimulates glioma cell migration, \\ invasion, colony formation, and tumor growth in mice \\ - High expression levels associated with poor \\ patient survival in LGG, HGG, and GBM \\ - Upregulated after TBI in humans and mice \\} & \makecell{\citep{sanden2015aberrant} \\\\ \citep{sanden2015aberrant}\\ \citep{barash2019heparanase}\\\\ 
\citep{barash2019heparanase,soni2017cd24}\\\\
\citep{li2014elevated}\\} \\
\hline
LAPTM5 & \makecell{- Negatively regulates T and B cell receptor signaling \\ - May be used by glioma to negatively \\ modulate antitumoral response of immune system \\ - High expression levels associated with poor survival in GBM \\ - Downregulated in other human cancers, such as neuroblastoma} & \makecell{\citep{ouchida2008lysosomal,ouchida2010role} \\ \citep{m2020aberrant}\\\\
\citep{m2020aberrant}\\
\citep{m2020aberrant,nuylan2016down}\\} \\
\hline
C1QB & \makecell{- Involved in initiation of classical complement pathway and mediation \\ of immunoregulatory functions in the prevention of autoimmunity\\ - Upregulated in diverse grade of gliomas \\ and plays important role in pathogenesis \\ - Involved in the conversion of PrPC into PrPSc by prion diseases \\ - Gene polymorphism associated with \\ SZ susceptibility in Armenian population \\ - Expression levels increase in association with \\ neurodegeneration in sporadic amyotrophic lateral sclerosis} & \makecell{\citep{thielens2017c1q} \\ \\ \citep{mangogna2019prognostic}\\\\ \citep{kanehisa2000kegg}\\ \citep{zakharyan2011association}\\\\ \citep{grewal1999c1qb}\\\\} \\
\hline
IL1B & \makecell{- Involved in mediation of inflammatory response, \\ cell proliferation, differentiation, and apoptosis \\ - Highly expressed in GBM and promotes glioma spheroid formation \\ - Involved in reduction of long-term potentiation in AD \\ - Active role in necroptosis, a programmed form of \\ necrosis involved in pathogenesis of neurological diseases \\ - Functions in NF-$\kappa$B signaling pathway \\ - Plays role in mitogen-activated protein kinase (MAPK) cascade which \\ functions in cell proliferation, differentiation and migration} & \makecell{\citep{o2016reference} \\\\ \citep{wang2019ccaat}\\
\citep{kanehisa2000kegg}\\
\citep{kanehisa2000kegg}\\ \\
\citep{kanehisa2000kegg}\\
\citep{kanehisa2000kegg}\\} \\
\hline
GPM6B & \makecell{- Involved in neuronal differentiation and myelination \\ - Highly expressed in GBM \\ - Transcriptional downregulation in hippocampus of depressed suicides \\ - Overexpressed in pre-B acute lymphoblastic leukemia and \\ detected in serum of patients with early stage ovarian cancer} & 
\makecell{\citep{dere2015gpm6b,werner2013critical} \\ \citep{castells2010development}\\
\citep{fuchsova2015altered}\\
\citep{charfi2014identification,urban2011overview}\\ \\
} \\
\hline
MEG3 & \makecell{- Suppresses proliferation, migration, and invasion of glioma cells\\ - Significantly downregulated in glioma \\ - Expression levels associated negatively with \\ tumor grading and positively with overall survival \\ - Upregulated in HD mouse model and after ischemia in adult mice \\ - Decreased expression levels in plasma of patients with TBI} & \makecell{\citep{qin2017long} \\ \citep{zhao2018long}\\
\citep{zhao2018long}\\\\
\citep{chanda2018altered,yan2016long}\\
\citep{shao2019research}\\} \\
\hline
SRGN & \makecell{- Involved in formation of mast cell secretory granules and storage \\ of proteases in mucosal mast cells and granzyme B in T-lymphocytes
 \\ - Active in granzyme-mediated apoptotic signaling pathway \\ - Preferentially secreted by cancer-associated \\ fibroblasts for promoting tumor growth in breast cancer \\ - Upregulated in HCC, colorectal cancer, primary \\ non-small cell lung carcinoma, and acute myeloid leukemia \\ - Promotes metastasis and associated with poor outcomes in these cancers} & \makecell{\citep{uniprot2019uniprot} \\ \\ \citep{uniprot2019uniprot} \\ \citep{tyan2012breast} \\ \\
\citep{he2013serglycin,xu2018srgn}
\\\citep{guo2017serglycin,peng2020lncrna} \\ \\} \\
\hline
CCL4L1 & \makecell{- Active in chemotaxis and inflammatory response \\ - Has redundant function as CCL4} & \makecell{\citep{uniprot2019uniprot} \\ \citep{howard2004functional}} \\
\hline
CCL4L2 & \makecell{- Active in chemotaxis and inflammatory response} & \makecell{\citep{uniprot2019uniprot}} \\
\hline
TUBA1A & \makecell{- Helps form and organize microtubules, necessary for cell division and \\ movement, and involved in neuronal migration in developing brain \\ - Active in the pathogenesis of AD and HD \\ - Commonly expressed across three MS related phenotypes \\ - Mutations cause lissencephaly with diffuse agyria \\ or pachygyria, associated with microcephaly, agenesis \\ of the corpus callosum, and cerebellar hypoplasia \\ - Up-regulated in breast cancer tumors} & \makecell{\citep{o2016reference} \\\\ \citep{kanehisa2000kegg}\\
\citep{mahurkar2013identification}\\
\citep{conti2017human,kato2015genotype}\\\\\\
\citep{nami2018genetics}\\} \\
\hline
DBNDD2 & \makecell{- Apoptosis response gene implicated in \\ neurodegeneration and neuronal injury \\ - Found in ubiquitinated lesions, including neurofibrillary \\ tangles and granulovacuolar degeneration bodies in AD} & \makecell{\citep{salinas2019whole,lucas2005human}\\\\ \citep{salinas2019whole,yin2006dysbindin}\\\\
} \\
\hline
C3 & \makecell{- Supports activation of all three pathways of complement activation \\ - Involved with neuroactive ligand-receptor interaction \\ - Plays role in the pathway for viral carcinogenesis \\ - Active in Kaposi sarcoma-associated herpesvirus infection pathway} & \makecell{\citep{sahu2001structure} \\ \citep{kanehisa2000kegg}\\ \citep{kanehisa2000kegg} \\ \citep{kanehisa2000kegg}} \\
\hline
FKBP5 & \makecell{- Functions in immunoregulation, \\ protein folding, and intracellular trafficking \\ - Inhibits proliferation and stimulates apoptosis of glioma cells \\ - Promotes tumor growth and chemoresistance in \\ other cancers through NF-$\kappa$B signaling regulation
 \\ - May be associated with structural changes of neural circuits related \\ to emotional control and mood regulation in major depressive disorder} & \makecell{\citep{o2016reference} \\\\ \citep{yang2015fk506}\\
\citep{yang2015fk506,romano2011fkbp51}\\\\
\citep{han2017influence}\\ \\
} \\
\hline
CSF1R & \makecell{- Controls production, differentiation, and function of macrophages \\ - Overexpressed in human HGG and GBM \\ - Inhibition blocks glioma progression, suppresses \\ tumor cell proliferation, and reduces tumor grade \\ - Plays role in MAPK cascade \\ - Involved in Ras signaling pathway, which regulates cell \\ proliferation, survival, growth, migration, and differentiation \\- Active in PI3K-Akt signaling pathway} & \makecell{\citep{o2016reference} \\ \citep{de2016csf1,bender2010sleeping}\\
\citep{yan2017inhibition}\\ \\
\citep{kanehisa2000kegg}\\
\citep{kanehisa2000kegg}\\ \\
\citep{kanehisa2000kegg}\\
} \\
\hline
PMP2 & \makecell{- Constituent of PNS and CNS myelin that stabilizes myelin \\ membranes and may play a role in lipid transport in Schwann cells \\ - Upregulated in astrocytoma \\ - Upregulated in melanoma cell lines and enhances cell invasion \\ - Mutation causes dominant demyelinating Charcot-Marie-Tooth \\ disease, a hereditary motor and sensory neuropathy of the PNS} & \makecell{\citep{o2016reference,uniprot2019uniprot} \\\\ \citep{cai2015screening}\\
\citep{graf2019myelin}\\
\citep{hong2016mutation}\\\\} \\
\hline
HLA-B & \makecell{- Helps immune system distinguish body's own \\ proteins from proteins made by foreign invaders \\ - Polymorphisms associated with independent MS \\ susceptibility in Japanese population and early-onset development \\ of myasthenia gravis, a chronic autoimmune, neuromuscular disease \\ - Significantly up-regulated in lung cancer patients \\ - Plays role in human T-cell leukemia virus 1 infection pathway \\ - Involved in EBV infection pathway \\ - Active in human papillomavirus infection pathway \\ - Active in Kaposi sarcoma-associated herpesvirus infection pathway \\ - Plays role in the pathway for viral carcinogenesis} & \makecell{\citep{uniprot2019uniprot} \\\\ \citep{ogawa2019next,varade2018novel}\\\\\\
\citep{liu2019three}\\
\citep{kanehisa2000kegg}\\
\citep{kanehisa2000kegg}\\
\citep{kanehisa2000kegg}\\
\citep{kanehisa2000kegg}\\
\citep{kanehisa2000kegg}\\
} \\
\hline
\end{longtable}
\end{center}
\setstretch{2}

\newpage

\bibliographystyle{vancouver-authoryear}
\bibliography{references}